%% file: main.tex
\makeatletter
\declare@file@substitution{revtex4-1.cls}{revtex4-2.cls}
\makeatother
\documentclass[twocolumn]{aastex63}
\pdfoutput=1 %
\usepackage{amsmath,amstext}
\usepackage{tikz} %
\usetikzlibrary{shapes.geometric, arrows}
\usepackage[T1]{fontenc}
\usepackage{apjfonts}
\usepackage{hyperref}
\usepackage[all]{hypcap}
\usepackage{url}
\usepackage{xspace}
\usepackage{color}
\usepackage{xcolor, colortbl}
\usepackage{savesym}
\savesymbol{tablenum}
\usepackage{siunitx}
\restoresymbol{SIX}{tablenum}
\usepackage{multirow}
\usepackage{tabularx}

\usepackage{placeins}
\usepackage{listings}
\usepackage{booktabs}

\usepackage{orcidlink}

\newcommand{\Spider}{\textsc{Spider}\xspace}
\newcommand{\spider}{\textsc{Spider}\xspace}
\newcommand{\Planck}{\textit{Planck}\xspace}
\newcommand{\planck}{\textit{Planck}\xspace}
\newcommand{\WMAP}{\textit{WMAP}\xspace}

\newcommand{\gdust}{\ensuremath{{P_\mathrm{d}}^+}\xspace}
\newcommand{\ldust}{\ensuremath{{P_\mathrm{d}}^-}\xspace}

\shorttitle{Analysis of Polarized Dust Emission Using \Spider Data}
\shortauthors{\Spider Collaboration}

\begin{document}

\title{Analysis of Polarized Dust Emission Using Data from the First Flight of \Spider}
\input spider_authors.tex

\begin{abstract}

\input{abstract}

\end{abstract}

\section{Introduction}

The polarization of the cosmic microwave background (CMB) contains a wealth of information about the contents, history, and origins of the Universe~\citep{S4science}. %
Observations of the $E$-mode (``gradient'', scalar, even-parity) polarization pattern reveal details of acoustic oscillations in the primordial plasma.
A measurement of a cosmological $B$-mode (``curl'', pseudoscalar, odd-parity) pattern on large angular scales would imply the presence of a background of primordial gravitational waves, and thus provide remarkable new insights into early Universe models \citep{Kamionkowski_review:2016, zaldarriaga_bmode}.

Precision studies of CMB polarization are complicated by polarized emission at similar frequencies from foreground sources within our own Galaxy, notably synchrotron radiation and thermal emission from dust particles \citep{dunkley_2009_fgforecast}.
Even at the highest Galactic latitudes, foreground emission is known to be far brighter than any primordial $B$-mode signal \citep{bicepplanck_2015}.
Observations at multiple frequencies can be used to differentiate CMB from foreground emission based upon their differing spectral energy distributions (SEDs) \citep{COBE_dust91,brandt1994}.
Several such component separation methods have been developed: see, \emph{e.g.}, \cite{Planck_CompSep} and references therein.
In order to stitch together such multi-frequency data, these methods must necessarily make assumptions about the nature of foregrounds: their spatial morphology, SED, or some combination thereof.
Component separation is thus an imperfect process, and there are always some uncertainties due to these assumptions \citep{remazeilles_2016, Hensley_2018}.

At frequencies above 100\,GHz, Galactic dust emission is the dominant polarized foreground on large angular scales \citep{planck2015_X}.%
Dust grains, heated by stellar radiation to ${\sim}20$~K, thermally re-emit that energy with an efficiency that depends on the physical characteristics of the grains.
Because emission is inefficient at
wavelengths much larger than the grain itself, dust emissivity is
expected to drop in the $mm$-wave regime where CMB measurements occur \citep{draine_2009}.
Aspherical grains that are ferro- or para-magnetic will preferentially align relative to the local magnetic field, leading to linearly-polarized emission with large-scale correlations \citep{purcell_dust}.
Motivated by this picture, it is common to model the dust SED\footnote{Note that, while Equation~\ref{eq:MBB} may be used to model both the intensity and polarization of dust emission, the same model parameters need not apply to both; indeed, \Planck has found a slightly steeper power law for polarization than for intensity in an all-sky analysis~\cite{Planck_2018_XI}.}
as a thermal blackbody at temperature $T_\mathrm{d}$, modified by a power-law emissivity with exponent $\beta_\mathrm{d}$:
\begin{equation}
\label{eq:MBB}
I(\nu) = \tau_0 \left( \frac{\nu}{\nu_0} \right)^{\beta_\mathrm{d}} B(\nu, T_\mathrm{d}).
\end{equation}
Here $B$ is the Planck spectrum for a blackbody emitter, $\nu_0$ is some reference frequency, and $\tau_0$ is the optical depth at the reference frequency.
Equation~\ref{eq:MBB} is referred to as a \emph{modified blackbody (MBB)}.

Though widely-used, this simple model is unlikely to reflect the full range of dust emission across the sky.
The detailed nature of dust grains remains a subject of ongoing research \citep{hensley2022astrodustpah, draine_2009}.
Their composition, size distribution, and density are all expected to vary from place to place, as are their radiative and magnetic environments.
Thus, we may naturally expect different dust populations in our Galaxy to exhibit different SEDs.
Such variation of the spatial morphology of a foreground component with observation frequency---or, equivalently, of the component SED across the sky---is often called \emph{frequency decorrelation}.
Furthermore, observations of the CMB constitute a collection of line-of-sight (LOS) integrals through the materials surrounding us, each of which may traverse heterogeneous populations.
In addition to complicating the dust SED, such \emph{LOS decorrelation} can yield variation in polarization angle with frequency even within a single map pixel \citep{tassis_2015}.

Decorrelation may significantly complicate efforts at component separation.
For example, \citet{poh_2017_losdust} have shown that naive extrapolation of dust templates made from 353\,GHz data to lower frequencies can bias the recovery of the tensor-to-scalar ratio $r \leq 1.5\times 10^{-3}$.
While an analysis by the BICEP team \citep{bicep_XVI} showed no evidence of dust decorrelation within their observing region, there is evidence for its presence in larger regions.
By identifying regions of the sky likely to contain magnetically misaligned clouds, \cite{pelgrims_2021} demonstrate that LOS frequency decorrelation is detectable within the \planck dataset, largely in the Northern Galactic cap.
\cite{ritacco} has further demonstrated LOS frequency decorrelation in large-scale regions close to the Galactic plane.
It is thus critically important to test the assumptions behind common foreground separation techniques through detailed studies of dust polarization.
Such studies will inform both the field of Galactic astrophysics as well as the prospects for foreground removal by future CMB experiments \citep{Hensley_2022, Hazumi_2020, S4science}.

This publication is part of a series describing results from the first flight of \Spider, a balloon-borne CMB telescope optimized to search for $B$-modes on degree angular scales.
These data from \spider provide a measurement of the angular power spectra of 4.8\% of the polarized sky at two frequencies: 95~and 150\,GHz \citep{bmode_paper}.
Dust emission is the dominant foreground at these at these frequencies.
Since the first \spider instrument did not include a channel optimized for dust, the analysis of these data relies heavily on \Planck 217 and 353\,GHz maps for component separation.
The published $B$-mode results from these data relied primarily on template subtraction using the XFaster power spectrum estimator (Section~\ref{sec:template_method}) alongside comparisons with alternate estimators and foreground-cleaning techniques (\emph{e.g.}, Section~\ref{subsec:SMICA}).

This paper exploits the sensitivity and sky coverage of the \spider data to explore the properties of the Galactic emission and to quantify the impact of the methods and models employed on component separation.
Section~\ref{sec:techniques} reviews both component separation techniques, and in Section~\ref{sec:spatial_variation} they are applied to two distinct subsets of the \spider region to probe for evidence of spatial variation in the nature of the diffuse polarized dust emission.
The fidelity of the MBB model is tested in a comparison of the results from the two sub-regions using two \Planck-derived dust templates (353-100\,GHz and 217-100\,GHz).
Section~\ref{sec:testing_modeling} examines different choices for modeling the dust SED within the full \spider region: the validity of the MBB model, the constancy of MBB parameters with angular scale, and the use of power-law models for the dust angular power spectrum.
Section~\ref{sec:los_dust} searches for LOS decorrelation by developing a custom estimator to look for evidence of variation of dust polarization angle with frequency in the \spider observing region.
Finally, Section~\ref{sec:pysm} compares the measured dust power spectrum for this sky region to templates generated by the PySM foreground modeling package.

\section{Component Separation Techniques}
\label{sec:techniques}

\input{techniques.tex}

\section{Spatial Dependence of Polarized Dust Emission Properties}
\label{sec:spatial_variation}

\input{spatial_variation.tex}

\section{Testing Standard Dust Modeling}
\label{sec:testing_modeling}

\input{testing_dust_modeling.tex}

\section{Line-of-sight Decorrelation of Dust}

\label{sec:los_dust}

\input{dust_angle.tex}

\section{Comparison to PySM Dust Models}\label{sec:pysm}

\input{comp_to_pysm.tex}

\section{Conclusion}

\input{conclusion}

\begin{acknowledgments}
\Spider is supported in the U.S. by the National Aeronautics and Space Administration under grants NNX07AL64G, NNX12AE95G, NNX17AC55G, and 80NSSC21K1986 issued through the Science Mission Directorate, and by the National Science Foundation through PLR-1043515.
Logistical support for the Antarctic deployment and operations was provided by the NSF through the U.S. Antarctic Program.
Support in Canada is provided by the Natural Sciences and Engineering Research Council and the Canadian Space Agency.
Support in Norway is provided by the Research Council of Norway.
The Dunlap Institute is funded through an endowment established by the David Dunlap family and the University of Toronto.
The Flatiron Institute is supported by the Simons Foundation.
JEG acknowledges support from the Swedish Research Council (Reg.\ no.\ 2019-03959) and the Swedish National Space Agency (SNSA/Rymdstyrelsen). This work is in part funded by the European Union (ERC, CMBeam, 101040169).
J.M.N.~acknowledges support from the Research Corporation for Science Advancement.
K.F. acknowledges support from DOE grant DE-SC0007859 at the University of Michigan.
We also wish to acknowledge the generous support of the David and Lucile Packard Foundation, which has been crucial to the success of the project.

The collaboration is grateful to the British Antarctic Survey, particularly Sam Burrell, for invaluable assistance with data and payload recovery after the 2015 flight.
We thank Brendan Crill and Tom Montroy for significant contributions to \Spider's development.
Some of the results in this paper have been derived using the HEALPix package \citep{HealPix}.
The computations described in this paper were performed on the GPC supercomputer at the SciNet HPC Consortium \citep{Scinet}.
SciNet is funded by the Canada Foundation for Innovation under the auspices of Compute Canada, the Government of Ontario, Ontario Research Fund - Research Excellence, and the University of Toronto.
\end{acknowledgments}

\clearpage
\appendix

\input{appendix_orion_superbubble}
\input{appendix_dustalpha}
\input{appendix_pysm}

\bibliographystyle{apj}
\bibliography{references}

\end{document}

%% file: spider_authors.tex
\newcommand\ANL{High Energy Physics Division, Argonne National Laboratory, Argonne, IL, USA 60439}
\newcommand\NU{Department of Physics \& Astronomy, Northwestern University, Evanston, IL 60208, USA}
\newcommand\CWRU{Physics Department, Case Western Reserve University, 10900 Euclid Ave, Rockefeller Building, Cleveland, OH 44106, USA}
\newcommand\Cardiff{School of Physics and Astronomy, Cardiff University, The Parade, Cardiff, CF24 3AA, UK}
\newcommand\UBC{Department of Physics and Astronomy, University of British Columbia, 6224 Agricultural Road,
Vancouver, BC V6T 1Z1, Canada}
\newcommand\Princeton{Department of Physics, Princeton University, Jadwin Hall, Princeton, NJ 08544, USA}
\newcommand\Caltech{Division of Physics, Mathematics and Astronomy, California Institute of Technology, MS 367-17, 1200 E. California Blvd., Pasadena, CA 91125, USA}
\newcommand\JPL{Jet Propulsion Laboratory, Pasadena, CA 91109, USA}
\newcommand\CITA{Canadian Institute for Theoretical Astrophysics, University of Toronto, 60 St. George Street, Toronto, ON M5S 3H8, Canada}
\newcommand\ASU{School Of Earth and Space Exploration, Arizona State University, 650 E Tyler Mall, Tempe, AZ 85281, USA}
\newcommand\UKZN{School of Mathematics, Statistics and Computer Science, University of KwaZulu-Natal, Durban, South Africa}
\newcommand\NITP{National Institute for Theoretical Physics (NITheP), KwaZulu-Natal, South Africa}
\newcommand\Imperial{Blackett Laboratory, Imperial College London, SW7 2AZ, London, UK}
\newcommand\Stockholm{The Oskar Klein Centre for Cosmoparticle Physics, Department of Physics, Stockholm University, AlbaNova, SE-106 91 Stockholm, Sweden}
\newcommand\Oslo{Institute of Theoretical Astrophysics, University of Oslo, P.O. Box 1029 Blindern, NO-0315 Oslo, Norway}
\newcommand\TorontoDunlap{Dunlap Institute for Astronomy and Astrophysics, University of Toronto, 50 St George Street, Toronto, ON M5S 3H4 Canada}
\newcommand\Toronto{Department of Astronomy and Astrophysics, University of Toronto, 50 St George Street, Toronto, ON M5S 3H4 Canada}
\newcommand\UIUCP{Department of Physics, University of Illinois Urbana-Champaign, 1110 W. Green Street, Urbana, IL 61801, USA}
\newcommand\NRAO{National Radio Astronomy Observatory, Charlottesville, NC 22903, USA}
\newcommand\Michigan{Department of Physics, University of Michigan, 450 Church Street, Ann  Arbor, MI 48109, USA}
\newcommand\TorontoP{Department of Physics, University of Toronto, 60 St George Street, Toronto, ON M5S 3H4 Canada}
\newcommand\Hopkins{Department of Physics and Astronomy, Johns Hopkins University, 3701 San Martin Drive, Baltimore, MD 21218 USA}
\newcommand\Goddard{NASA Goddard Space Flight Center, Code 665, Greenbelt, MD 20771, USA}
\newcommand\APC{APC, Univ. Paris Diderot, CNRS/IN2P3, CEA/Irfu, Obs de Paris, Sorbonne Paris Cit\'e, France}
\newcommand\PennState{Department of Astronomy and Astrophysics, Pennsylvania State University, 520 Davey Lab, University Park, PA 16802, USA}
\newcommand\NIST{National Institute of Standards and Technology, 325 Broadway Mailcode 817.03, Boulder, CO 80305, USA}
\newcommand\Stanford{Department of Physics, Stanford University, 382 Via Pueblo Mall, Stanford, CA 94305, USA}
\newcommand\SLAC{ SLAC National Accelerator Laboratory, 2575 Sand Hill Road, Menlo Park, CA 94025, USA}
\newcommand\PrincetonEngineering{Department of Mechanical and Aerospace Engineering, Princeton University, Engineering Quadrangle, Princeton, NJ 08544, USA}
\newcommand\Fermilab{Fermi National Accelerator Laboratory, P.O. Box 500, Batavia, IL 60510-5011, USA}
\newcommand\KICPChicago{Kavli Institute for Cosmological Physics, University of Chicago, 5640 S Ellis Avenue, Chicago, IL 60637 USA}
\newcommand\AAUChicago{Department of Astronomy and Astrophysics, University of Chicago, 5640 S Ellis Avenue, Chicago, IL 60637 USA}
\newcommand\Orsay{Institut d'Astrophysique Spatiale, Orsay, France}
\newcommand\MPI{Max-Planck-Institute for Astronomy, Konigstuhl 17, 69117, Heidelberg, Germany}
\newcommand\LAIM{Laboratoire AIM, Paris-Saclay, CEA/IRFU/SAp - CNRS - Universit\'e Paris Diderot, 91191, Gif-sur-Yvette Cedex, France}
\newcommand\WUSTL{Department of Physics, Washington University in St. Louis, 1 Brookings Drive, St.  Louis, MO 63130, USA}
\newcommand\MCSS{McDonnell Center for the Space Sciences, Washington University in St. Louis, 1 Brookings Drive, St.  Louis, MO 63130, USA}
\newcommand\Austin{Department of Physics, University of Texas, 2515 Speedway, C1600, Austin, TX 78712, USA}
\newcommand\Weinberg{Weinberg Institute for Theoretical Physics, Texas Center for Cosmology and Astroparticle Physics, Austin, TX 78712, USA}
\newcommand\McGill{Department of Physics, McGill University, 3600 Rue University, Montreal, QC, H3A 2T8, Canada}
\newcommand\StewardObs{Steward Observatory, 933 North Cherry Avenue, Tucson, AZ, 85721, USA}
\newcommand\Iceland{Science Institute, University of Iceland, 107 Reykjavik, Iceland}
\newcommand\Flatiron{Center for Computational Astrophysics, Flatiron Institute, New York, NY 10010, USA}

\author{\spider Collaboration}
\noaffiliation{}

\author{ P.~A.~R.~Ade }
\affiliation{\Cardiff}

\author{ M.~Amiri }
\affiliation{\UBC}

\author{ S.~J.~Benton \orcidlink{0000-0002-4214-9298}}
\affiliation{\Princeton}

\author{ A.~S.~Bergman }
\affiliation{\Princeton}

\author{ R.~Bihary }
\affiliation{\CWRU}

\author{ J.~J.~Bock }
\affiliation{\Caltech}
\affiliation{\JPL}

\author{ J.~R.~Bond }
\affiliation{\CITA}

\author{ J.~A.~Bonetti }
\affiliation{\JPL}

\author{ S.~A.~Bryan }
\affiliation{\ASU}

\author{ H.~C.~Chiang }
\affiliation{\McGill}
\affiliation{\UKZN}

\author{ C.~R.~Contaldi }
\affiliation{\Imperial}

\author{ O.~Dor{\'e} }
\affiliation{\Caltech}
\affiliation{\JPL}

\author{ A.~J.~Duivenvoorden \orcidlink{0000-0003-2856-2382}}
\affiliation{\Flatiron}
\affiliation{\Princeton}

\author{ H.~K.~Eriksen }
\affiliation{\Oslo}

\author{ J.~P.~Filippini \orcidlink{0000-0001-8217-6832}}
\affiliation{\UIUCP}

\author{ A.~A.~Fraisse }
\affiliation{\Princeton}

\author{ K.~Freese }
\affiliation{\Austin}
\affiliation{\Weinberg}
\affiliation{\Stockholm}

\author{ M.~Galloway }
\affiliation{\Oslo}

\author{ A.~E.~Gambrel }
\affiliation{\KICPChicago}

\author{ N.~N.~Gandilo }
\affiliation{\StewardObs}

\author{ K.~Ganga }
\affiliation{\APC}

\author{ S.~Gourapura \orcidlink{0000-0002-8149-0632}}
\affiliation{\Princeton}

\author{ R.~Gualtieri \orcidlink{0000-0003-4245-2315}}
\affiliation{\UIUCP}
\affiliation{\NU}

\author{ J.~E.~Gudmundsson \orcidlink{0000-0003-1760-0355}}
\affiliation{\Iceland}
\affiliation{\Stockholm}

\author{ M.~Halpern }
\affiliation{\UBC}

\author{ J.~Hartley  \orcidlink{0009-0006-9861-9718}}
\affiliation{\TorontoP}

\author{ M.~Hasselfield }
\affiliation{\PennState}

\author{ G.~Hilton }
\affiliation{\NIST}

\author{ W.~Holmes }
\affiliation{\JPL}

\author{ V.~V.~Hristov }
\affiliation{\Caltech}

\author{ Z.~Huang }
\affiliation{\CITA}

\author{ K.~D.~Irwin }
\affiliation{\Stanford}
\affiliation{\SLAC}

\author{ W.~C.~Jones \orcidlink{0000-0002-3636-1241}}
\affiliation{\Princeton}

\author{ A.~Karakci }
\affiliation{\Oslo}

\author{ C.~L.~Kuo }
\affiliation{\Stanford}

\author{ Z.~D.~Kermish }
\affiliation{\Princeton}

\author{ J.~S.-Y.~Leung }
\affiliation{\Toronto}
\affiliation{\TorontoDunlap}

\author{ S.~Li \orcidlink{0000-0002-8896-911X}}
\affiliation{\Princeton}
\affiliation{\PrincetonEngineering}

\author{D.~S.~Y.~Mak}
\affiliation{\Imperial}

\author{ P.~V.~Mason }
\affiliation{\Caltech}

\author{ K.~Megerian }
\affiliation{\JPL}

\author{ L.~Moncelsi }
\affiliation{\Caltech}

\author{ T.~A.~Morford }
\affiliation{\Caltech}

\author{ J.~M.~Nagy \orcidlink{0000-0002-2036-7008}}
\affiliation{\CWRU}

\author{ C.~B.~Netterfield }
\affiliation{\Toronto}
\affiliation{\TorontoP}

\author{ M.~Nolta }
\affiliation{\CITA}

\author{ R. O\rq Brient}
\affiliation{\JPL}

\author{ B.~Osherson }
\affiliation{\UIUCP}

\author{ I.~L.~Padilla \orcidlink{0000-0002-0024-2662}}
\affiliation{\CWRU}
\affiliation{\Toronto}
\affiliation{\Hopkins}

\author{ B.~Racine }
\affiliation{\Oslo}

\author{ A.~S.~Rahlin \orcidlink{0000-0003-3953-1776}}
\affiliation{\AAUChicago}
\affiliation{\KICPChicago}

\author{ C.~Reintsema }
\affiliation{\NIST}

\author{ J.~E.~Ruhl \orcidlink{0000-0001-5875-4751}}
\affiliation{\CWRU}

\author{ M.~C.~Runyan }
\affiliation{\Caltech}

\author{ T.~M.~Ruud }
\affiliation{\Oslo}

\author{ J.~A.~Shariff }
\affiliation{\CITA}

\author{ E.~C.~Shaw \orcidlink{0000-0001-5644-8750}}
\affiliation{\UIUCP}
\affiliation{\Austin}
\affiliation{\Weinberg}

\author{ C.~Shiu \orcidlink{0000-0002-6635-5950} }
\affiliation{\Princeton}

\author{ J.~D.~Soler }
\affiliation{\MPI}

\author{ X.~Song }
\affiliation{\Princeton}

\author{ A.~Trangsrud }
\affiliation{\Caltech}
\affiliation{\JPL}

\author{ C.~Tucker }
\affiliation{\Cardiff}

\author{ R.~S.~Tucker }
\affiliation{\Caltech}

\author{ A.~D.~Turner }
\affiliation{\JPL}

\author{ J.~F.~van~der~List }
\affiliation{\Princeton}

\author{ A.~C.~Weber }
\affiliation{\JPL}

\author{ I.~K.~Wehus }
\affiliation{\Oslo}

\author{ D.~V.~Wiebe }
\affiliation{\UBC}

\author{ E.~Y.~Young }
\affiliation{\Stanford}
\affiliation{\SLAC}

\correspondingauthor{Jeffrey~P.~Filippini}
\email{jpf@illinois.edu}

%% file: abstract.tex
Using data from the first flight of \Spider and from \Planck HFI, we
probe the properties of polarized emission from interstellar dust in
the \Spider observing region.  Component separation algorithms
operating in both the spatial and harmonic domains are applied to
probe their consistency and to quantify modeling errors associated
with their assumptions.  Analyses of diffuse Galactic dust emission 
spanning the full \Spider region
demonstrate \emph{i)} a spectral energy distribution that is broadly consistent with a modified-blackbody
(MBB) model with a spectral index of $\beta_\mathrm{d}=1.45\pm0.05$
$(1.47\pm0.06)$ for {$E$ ($B$)-mode} polarization, slightly lower than
that reported by \Planck for the full sky; \emph{ii)} an angular
power spectrum broadly consistent with a power law; and \emph{iii)}
no significant detection of line-of-sight polarization decorrelation.  
Tests of several
modeling uncertainties find only a modest impact
($\sim$10\% in $\sigma_r$) on 
\spider's sensitivity to the cosmological tensor-to-scalar ratio.
The size of the \Spider region
further allows for a statistically meaningful analysis of the
variation in foreground properties within it.  Assuming a fixed dust
temperature $T_\mathrm{d}=19.6$\,K, an analysis of two independent
sub-regions of that field results in inferred values of
$\beta_\mathrm{d}=1.52\pm0.06$ and $\beta_\mathrm{d}=1.09\pm0.09$,
which are inconsistent at the $3.9\,\sigma$ level.
Furthermore, a joint analysis of \Spider and \Planck 217 and 353\,GHz
data within one sub-region is inconsistent with a
simple MBB at more than $3\,\sigma$, assuming a common morphology of
polarized dust emission over the full range of frequencies.  This evidence of variation may inform the component-separation approaches of future CMB polarization experiments.

%% file: techniques.tex
As in~\citet{bmode_paper}, our analyses in this paper rely on two main
component separation techniques: foreground-template subtraction, and
Spectral Matching Independent Component Analysis (SMICA).
In this section, we present baseline versions of these methods.
In Sections~\ref{sec:spatial_variation}
and~\ref{sec:testing_modeling} we explore variations in the
assumptions presented below in order to test our understanding of the
polarized emission from interstellar dust.

\subsection{Template Subtraction}
\label{sec:template_method}

In the baseline cosmological analysis of the \Spider first flight
data \citep{bmode_paper}, we remove polarized dust emission from
the \Spider 95~and 150\,GHz maps by subtracting from them a scaled
template for this emission derived from \Planck data.\footnote{In both
the analyses presented in \citet{bmode_paper} and in this
publication, we use release 3.01 of the \planck
data \citep{planck18_hfi} unless otherwise noted.}
The \spider implementation of foreground-template
subtraction follows the approach pioneered
by \citet{wmap_threeyear} in the analysis of the \WMAP polarization
data.
Within any given map pixel observed at frequency $\nu$, the measured
Stokes parameters, $S_\nu$, are each modeled as the sum of three
components: CMB, dust, and noise:
\begin{equation}
  \label{eq:dust1}
  S_\nu = S^{\,\mathrm{CMB}} + A_{\nu,\nu_0} S_{\nu_0}^{\,\mathrm{dust}} + n_\nu\,.
\end{equation}
The uniform (pixel-independent) scaling factor $A_{\nu,\nu_0}$ connects the dust map
at the observing frequency $\nu$ to that at the reference frequency
$\nu_0=353$\,GHz.

We construct our dust template by subtracting the \Planck 100\,GHz map from the 353\,GHz map:
$S_{\nu_0}^{\,\mathrm{t}}\equiv S_{\nu_0}-S_{100}$.
Since the CMB component is independent of frequency in these units,
this removes the CMB signal at the cost of a modest increase in noise.
To produce a cleaned CMB map, we scale this dust template by a fitting parameter $\alpha$
and subtract it from $S_{\nu}$, giving
\begin{eqnarray}
\nonumber
S_\nu - \alpha\,S_{\nu_0}^{\,\mathrm{t}} & = & S^{\,\mathrm{CMB}} +
(A_{\nu,\nu_0} - \alpha\,[1-A_{100,\nu_0}])\,S_{\nu_0}^{\,\mathrm{dust}} \\
 & + & n_\nu - \alpha\,n_{\nu_0}^{\,\mathrm{t}}.
\label{eqn:alpha}
\end{eqnarray}
Here $n_\nu$ is the noise component at the frequency $\nu$, and
$n_{\nu_0}^{\,\mathrm{t}}\equiv n_{\nu_0}-n_{100}$.

Following \citet{bmode_paper}, we use the XFaster
algorithm \citep{Gambrel_2021} to fit for values of $\alpha=A_{\nu,\nu_0}/(1-A_{100,\nu_0})$
and the tensor-to-scalar ratio, $r$, in a simultaneous fit to the
\spider $EE$ and $BB$ spectra.
Note that, before
subtraction, the dust template is ``reobserved,'' \emph{i.e.}, injected as
an input sky through the full \spider simulation pipeline; it is thus
subjected to the same scan strategy, beam smoothing, time-domain
filtering, and map-making as the sky observed by \spider.
The reobserved template, like the \spider data, is corrected in map space
for the temperature-to-polarization leakage induced by the pipeline.

The template method implements a single scalar-valued factor $\alpha$
for each \Spider frequency multiplying the entire map.
Thus, this method assumes that the morphology of the polarized dust emission
is independent of frequency (at least between 353\,GHz  and 100\,GHz) or, equivalently,
that the dust SED is constant across the relevant sky area.
Note, however, that this method does \emph{not} impose assumptions on the dust SED itself.
We will revisit this assumption in our analysis below, notably in
Sections~\ref{sec:spatial_variation} and~\ref{sec:los_dust}.

\subsection{SMICA}\label{subsec:SMICA}

Spectral Matching Independent Component Analysis \citep[SMICA;][]{SMICA}
is a method for constructing linear combinations among a collection of maps in order
to isolate desired sky components (\emph{e.g.}, dust or CMB).
In contrast to the map-space template analysis described above, SMICA operates in
multipole space, acting on the spherical harmonic coefficients, $a_{\ell m}$, for each map.
Given a collection of
$n$ input frequency maps,\footnote{Throughout this paper, $n=6$: for each polarization mode we use maps at
all four polarized \Planck~HFI frequencies (100, 143, 217, and 353\,GHz) and
the two \Spider frequencies (95 and 150\,GHz).} we construct
$\pmb{Y} = (y_1, \cdots, y_n)$, where each $y_i$ is a row vector of
$a_{\ell m}$ coefficients for a single map $i$.
We use the SMICA algorithm to construct the column vector
of map weights, $w$, such that the linear combination
$s\equiv w^T \pmb{Y}$ is an estimate of the spherical harmonic coefficients for our desired sky component.
Throughout this section, lower case letters denote vectors, and upper case bold letters denote matrices.

The weights are optimized with respect to some model of the assumed sky components.
Given a column vector, $a$, containing the frequency scaling of a given sky component across
the various maps (uniform for CMB, MBB for dust), the multipole-binned weights, $w_b$, for this sky component are given by
\begin{equation}
  \label{eq:SMICA_weights}
  w_b^T =
        \left(a^T \widetilde{\pmb{R}}_b^{-1} a \right)^{-1} a^T \widetilde{\pmb{R}}_b^{-1},
\end{equation}
where $\widetilde{\pmb{R}}_b$ is the binned modeled covariance matrix containing CMB, dust, and noise components. These weights minimize the variance of the reconstructed harmonic map, $\mathrm{var(s)} = w_b^T \widetilde{\pmb{R}}_b w_b$ under the
signal-preserving constraint $w_b^T a = 1$  \citep{Hurier_2013}.

When reconstructing component maps, each harmonic mode $a_{\ell m}$ is scaled by the corresponding $w_b$ for its appropriate bin. The reconstructed maps, $S$, are the inverse spherical harmonic transform of these weighted collections of harmonic modes.

SMICA's power and flexibility lie in the freedom to impose (or relax) different assumptions about the components in this model, and thus about the relationships among the weights.
As discussed in detail in \citet{bmode_paper}, the SMICA implementation for
the \Spider $B$-mode analysis assumed a single dust component with a MBB SED.
In this analysis we make two improvements to SMICA dust modeling: we do not impose an SED on
the dust component, and we improve the power-spectrum estimation for the dust through a more accurate transfer function. Both changes are described further below.

In this analysis we do
\emph{not} assume a MBB dust SED, but instead fit for the dust $a$ as for any other parameter of the
model defining $\widetilde{\pmb{R}}_b$ (which includes the CMB, dust
emission, and noise) using the same maximum-likelihood fitting
procedure as in \citet{bmode_paper}.
We choose the arbitrary
overall normalization of $a$ such that its 353\,GHz element is set to 1.
Fitting for the dust $a$ in this manner effectively trades precision for accuracy:
the frequency-dependence of the reconstructed dust component will not be biased by any
assumptions about the dust SED, but the uncertainty on the
other fitted parameters (\emph{e.g.}, the amplitude of the emission) will
likely be larger; it could be smaller, but only if the SED we would otherwise
assume were an especially poor description of the data.

In previous analyses \citep{bmode_paper}, the sky components of $\widetilde{\pmb{R}}_b$ are corrected for effects from filtering, beam smoothing, and mode-coupling kernels through a $J_{b b'}$ transfer matrix, as per \cite{Leung_2022}.
While this approach is accurate for CMB spectra, the different angular power spectrum of dust make it inaccurate for an unbiased recovery of the full-sky dust spectrum, especially at larger scales.
In this paper we modify the model to apply $J_{b b'}$ \emph{only} to the CMB component, and instead apply a diagonal $F_b$ to the foreground component.
$F_b$ is the ratio of the output to input power for each bin and is constructed from an ensemble of reobserved Gaussian dust realizations.
These realizations follow a power-law scaling of amplitude with angular scale, with an exponent $\gamma = -2.30$, pivoting around $\ell = 80$.

\begin{figure}
   \centering
   \includegraphics[width=\columnwidth]{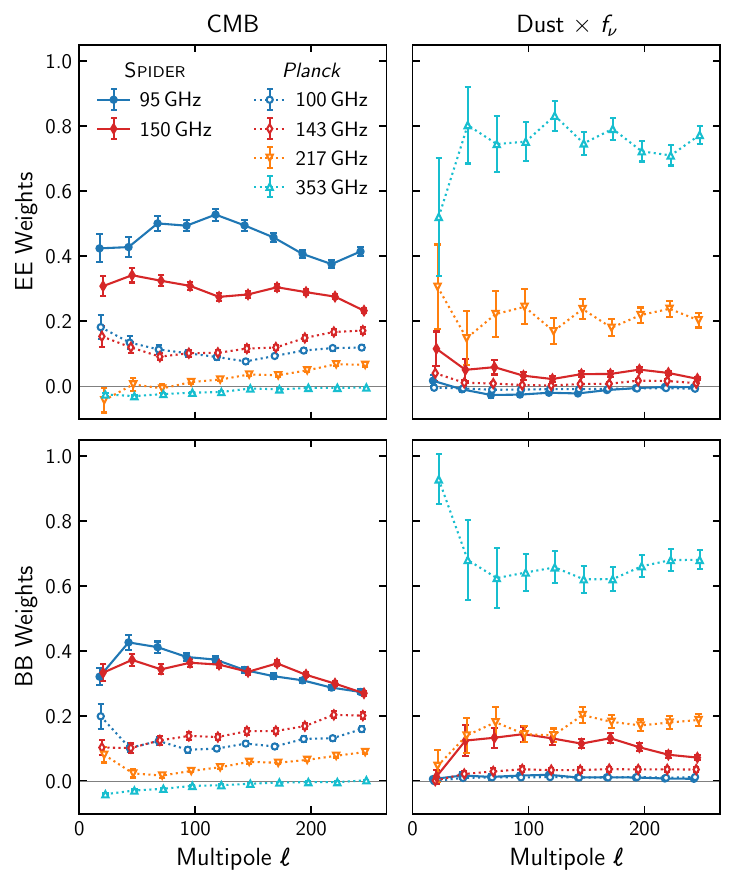}
   \caption{SMICA derived $EE$ (\textit{top}) and $BB$ (\textit{bottom}) weights
     to recover cleaned CMB (\textit{left}) and dust (\textit{right})
     component $Q$/$U$ maps.  The \spider 95 and 150\,GHz maps contribute
     predominantly toward the CMB signal, recovered (dominant positive weights)
     in the left panels and removed (subdominant and/or negative weights) in the
     right panels.  The \planck 217 and 353\,GHz maps contribute predominantly
     toward the dust signal, \textit{removed} in the left panels and
     \textit{recovered} in the right panels.  The dust weights are scaled with
     the fitted dust amplitude $f_\nu$ to indicate their signal contribution.
     \label{fig:smica_all_weights}}
\end{figure}

\begin{figure}
\centering
\includegraphics[width=\columnwidth]{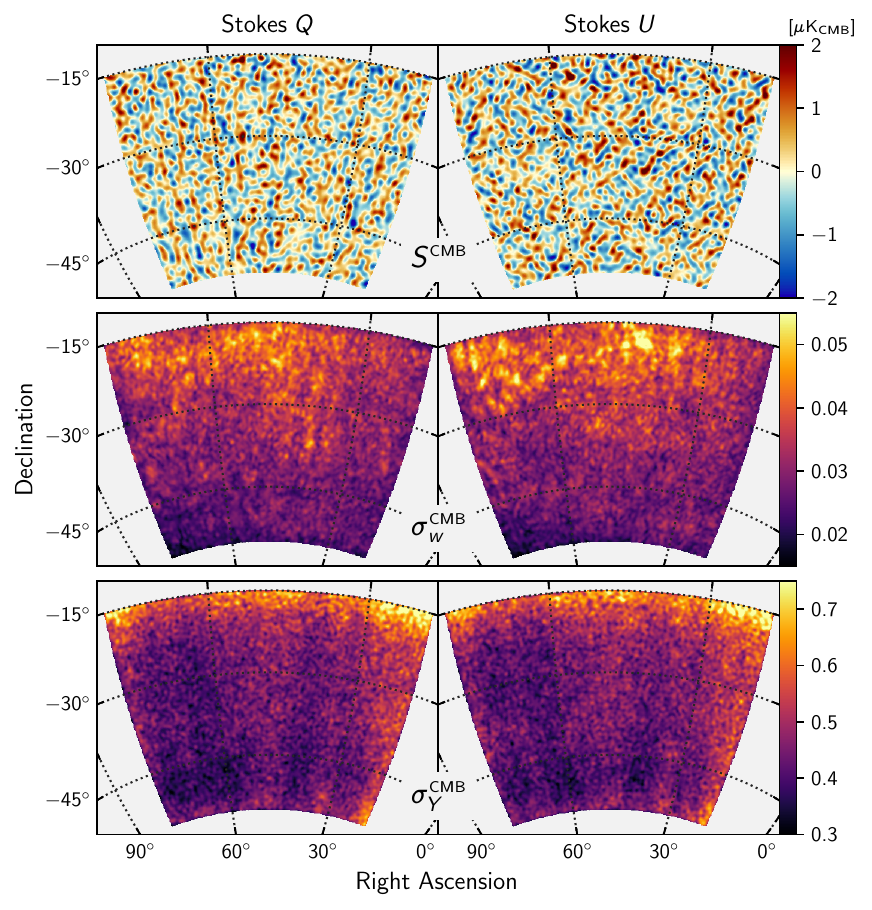}
\caption{(\textit{top}) SMICA derived component separated Stokes $Q$/$U$ CMB maps
computed with the CMB weights shown in the left panels of
Figure~\ref{fig:smica_all_weights}. (\textit{middle}) The contribution
$\sigma_w$ to the uncertainty in the CMB maps due to uncertainty in the SMICA
weights. (\textit{bottom}) The contribution $\sigma_Y$ to the CMB map uncertainty
due to the uncertainty in the input \spider and \planck data maps. All maps are
in units of $\mu$K$_\mathrm{CMB}$. \label{fig:smica_cmb_maps}}
\end{figure}

\begin{figure}
\centering
\includegraphics[width=\columnwidth]{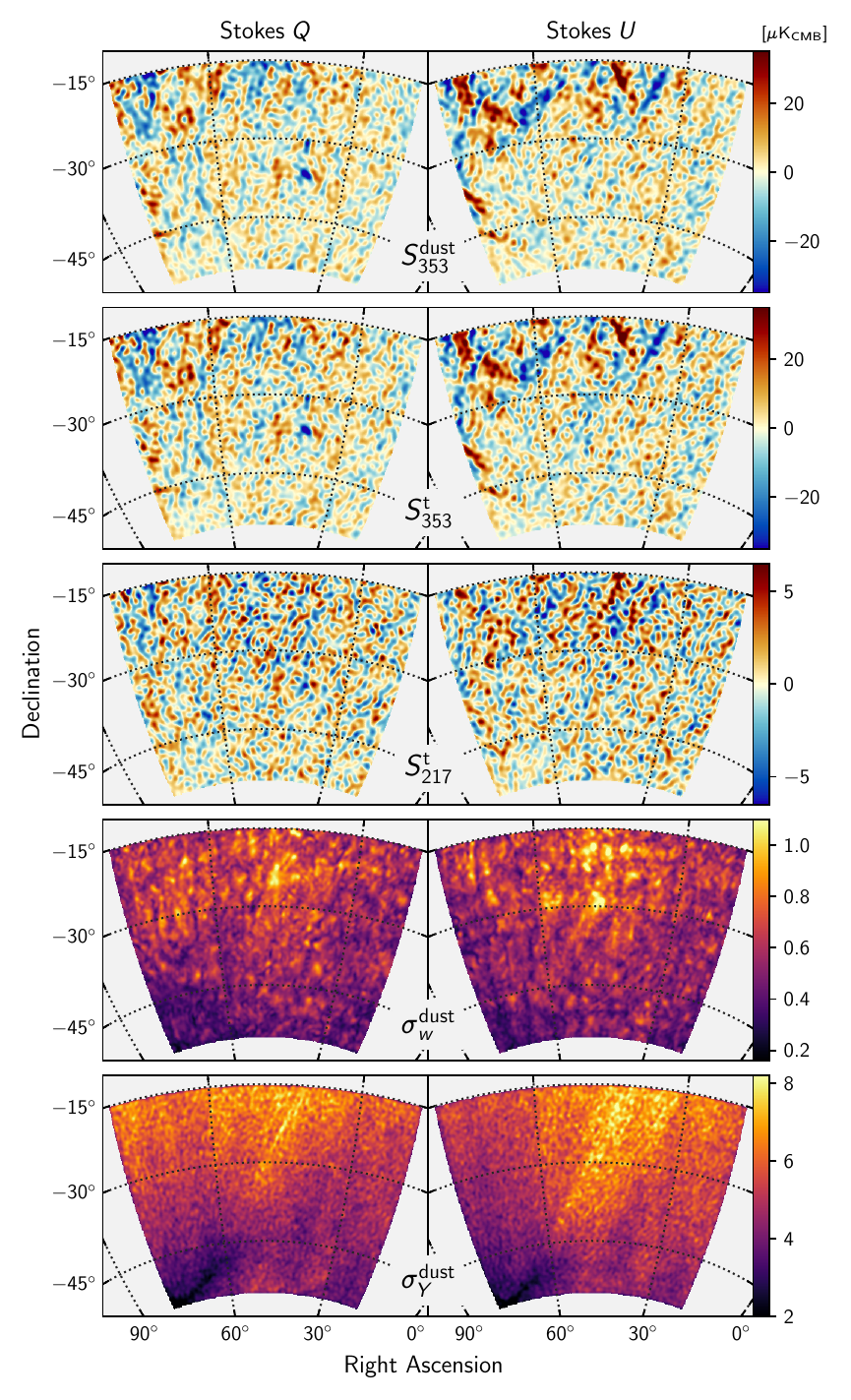}
\caption{(\textit{row 1}) SMICA derived component separated Stokes $Q$/$U$ dust
maps, computed with the dust weights shown in the right panels of
Figure~\ref{fig:smica_all_weights}, and scaled to the 353\,GHz reference
frequency. (\textit{rows 2 and 3}) The 353\,GHz and 217\,GHz \planck map
templates are reobserved and band-limited to $\ell \in [8, 258]$.  The SMICA dust map shows
structure that is most consistent with the 353\,GHz template, the predominant
contribution to the component weights. (\textit{row 4}) The contribution
$\sigma_w$ to the uncertainty in the SMICA dust maps due to uncertainty in the
weights. (\textit{row 5}) The contribution $\sigma_Y$ to the dust map
uncertainty due to the uncertainty in the input \spider and \planck data maps.
Note that the morphology of $\sigma_Y$ is inconsistent with that of
$\sigma_w$. All maps are in units of $\mu$K$_\mathrm{CMB}$.
\label{fig:smica_dust_maps}}
\end{figure}

Figure~\ref{fig:smica_all_weights} shows
the weights by which each \spider and \Planck~HFI frequency map contribute
to the component maps computed by SMICA. For the dust component, the
relative contribution of the map at frequency $\nu$ is given by its
SMICA ($a_{\ell m}$) weight multiplied by
$f_\nu\equiv a_\nu^\mathrm{dust}$, the $\nu$-component of the vector
$a^\mathrm{dust}$, the dust frequency scaling vector fitted to the
data by SMICA. The main contributor to both the $E$-mode and $B$-mode
dust signals in Figure~\ref{fig:smica_all_weights} is the
353\,GHz \Planck map, which has the highest ratio of
dust signal to noise. In the $B$-mode dust reconstruction, all
contributions are positive, as expected in the absence of CMB $B$-mode
power.

For the CMB, the SMICA weight for a given frequency map directly represents its relative contribution, since
$a^\mathrm{CMB}$ is the identity vector. The \spider maps provide the dominant
contribution to both the $E$ and $B$-modes, reflecting the higher ratios of CMB signal to noise in
these maps than in any of the \Planck HFI maps.

The $Q$ and $U$ maps
derived by SMICA, along with their estimated uncertainties,
are shown in Figures~\ref{fig:smica_cmb_maps} and~\ref{fig:smica_dust_maps} for the CMB and dust components, respectively.
The covariance $\sigma_{w}$ of the weights and the noise covariance $\sigma_Y$ contribute in quadrature to the total per-pixel uncertainty on the estimates of the component maps $S$.
The former represents the uncertainty due to signal reconstruction, while the latter (the dominant term) represents the per-pixel noise after a weighted average.
Both are presented in separate panels of Figures~\ref{fig:smica_cmb_maps} and~\ref{fig:smica_dust_maps}.

In Figure~\ref{fig:smica_dust_maps}, the morphology of $\sigma_Y^\textrm{dust}$ is similar to noise estimates for \planck 353\,GHz, as that map dominates the weight.
The uncertainty on the dust reconstruction, $\sigma_w^{\textrm{dust}}$, is morphologically distinct from the noise.
In particular, the larger uncertainty in the upper third of the $\sigma_w^{\textrm{dust}}$ maps does not trace the reconstructed dust signal strength or the noise.
This unexpected structure could come from
unidentified systematic errors in the input \Planck or \spider maps,
although evidence for the latter was not found in the expansive
suite of null tests published in \citet{bmode_paper}. A physical
origin is also possible: frequency decorrelation, in which dust
emission at a given frequency is not well correlated with that at
other frequencies, would prevent an accurate reconstruction of this
emission by SMICA, as the formalism here
assumes that a single frequency scaling factor applies to all
pixels. Given a specific frequency decorrelation model, \emph{e.g.}, a
mixture of two dust populations, each characterized by its own
spectral index, simulations could be run through SMICA in an attempt
to reproduce the structure observed in the dust uncertainty maps. Such
a study is beyond the scope of our analysis.

Figure~\ref{fig:smica_dust_maps} also compares the SMICA dust component map $S\mathrm{_{353}^{dust}}$ to the \planck dust template maps $S\mathrm{_{\nu_{0}}^{t}}$ constructed with $\nu_0 =$~353\,GHz and $\nu_0 =$~217\,GHz.
The template maps have been band-limited to $\ell \in [8, 258]$, to match the scales for which SMICA weights are fit.
The $S\mathrm{_{353}^{dust}}$ and $S\mathrm{_{353}^{t}}$ maps look very similar, which is not surprising since both are dominated by \planck 353\,GHz data.
$S\mathrm{_{217}^{t}}$ has much lower dust amplitude than $S\mathrm{_{353}^{t}}$, and lower signal-to-noise ratio.
The visible structures in these two template maps are similar but not identical.

%% file: spatial_variation.tex
In our baseline cosmological analysis \citep{bmode_paper}, we use
foreground-template subtraction (Section~\ref{sec:template_method}) to
subtract the polarized emission from interstellar dust from the
\spider maps. As reflected in the scalar nature of $A_{\nu,\nu_0}$
(and therefore of~$\alpha$) in Equation~\ref{eqn:alpha}, this
procedure assumes that the morphology of this emission
is independent of frequency. This assumption could
easily be inadequate, \emph{e.g.}, if the temperature of dust grains
changes across the \spider region, or if two or more dust
populations---even if they are well mixed---permeate it. In
Section~\ref{sec:spatial_variation_like}, we perform
foreground-template subtraction independently in two non-overlapping,
physically motivated subregions of the \spider field
(described in Section~\ref{sec:generating_subregions}), and assess the consistency
of the $\alpha$ values derived in the two subregions. We discuss the physical interpretation of these results in Section~\ref{sec:spatial_variation_betad_Td}.

\subsection{Delineating Subregions}
\label{sec:generating_subregions}

\begin{figure}
  \centering
  \includegraphics[width=0.95\columnwidth]{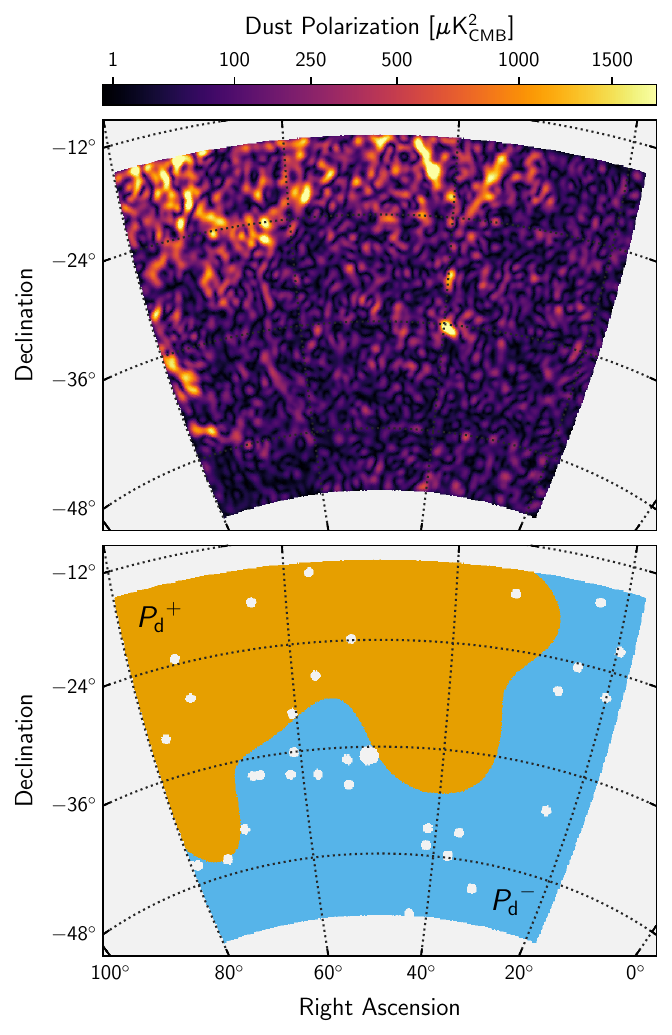}
  \caption{(\textit{top}) The polarization $P^2 = Q^2 + U^2$ of the
    SMICA-derived dust component (first row of Figure~\ref{fig:smica_dust_maps}).
    (\textit{bottom}) Mask regions \gdust and \ldust, containing more
    or less polarized dust, respectively, as inferred from the
    polarization map. \label{fig:smica_masks}}
\end{figure}

We define two physically motivated subregions within the \spider field based upon a SMICA-generated map of 
the total polarized dust emission power, $P^2=Q^2+U^2$ (Figure~\ref{fig:smica_masks}, \emph{top}).
This map is constructed from the $Q\mathrm{_{353}^{dust}}$ and $U\mathrm{_{353}^{dust}}$ Stokes
parameter maps described in Section~\ref{subsec:SMICA} and shown in Figure~\ref{fig:smica_dust_maps}.
After smoothing this map with a Gaussian of $8^\circ$~FWHM, we separate the map into two complementary 
subregions along the $P^2 = 100\,\mu \mathrm{K}_\mathrm{CMB}^2$ contour   (Figure~\ref{fig:smica_masks}, \emph{bottom}).
We denote by \gdust (\ldust) the subregion in which $P^2$ is greater
(less) than this threshold. 
This procedure produces two simply-connected subregions (prior to masking point sources) that are separated
by a smooth boundary, features that facilitate power spectrum estimation \citep{Gambrel_2021}.
After point-source masking according to
the procedure described in \citet{bmode_paper}, \gdust and \ldust cover roughly equal
sky fractions of 2.4\% and 2.3\%, respectively.
We note in passing that \ldust
overlaps with the ``Southern Hole,'' a region of the
Southern sky that has long been known to exhibit low \emph{total} dust
emission, and that has more recently been mapped with
high \emph{polarization} sensitivity
\citep{BK18_2021, balkenhol2023measurement}.

\subsection{Template Subtraction by Subregion: Evidence for Spatial~Variation}
\label{sec:spatial_variation_like}

\begin{figure*}
  \centering
  \includegraphics[width=\columnwidth]{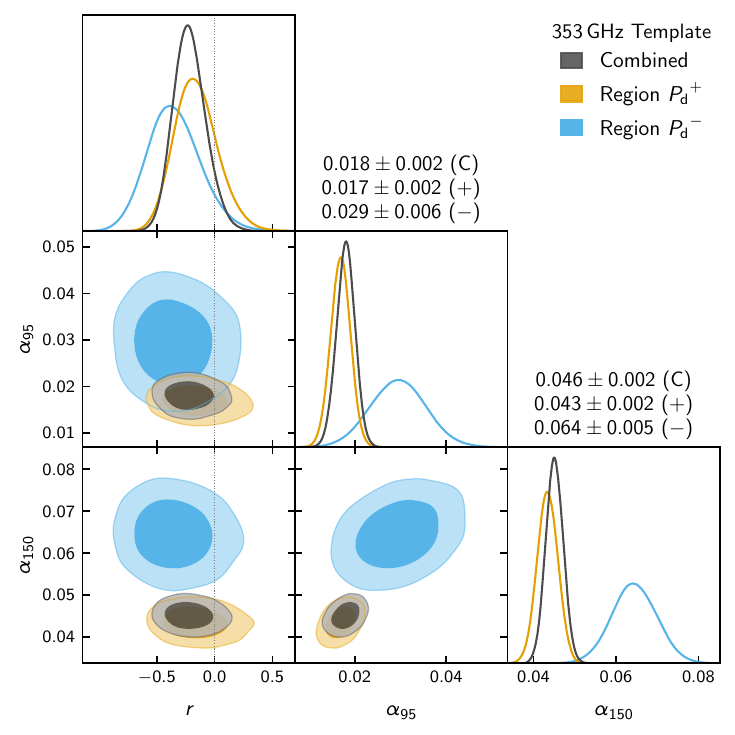}
  \hfill
  \includegraphics[width=\columnwidth]{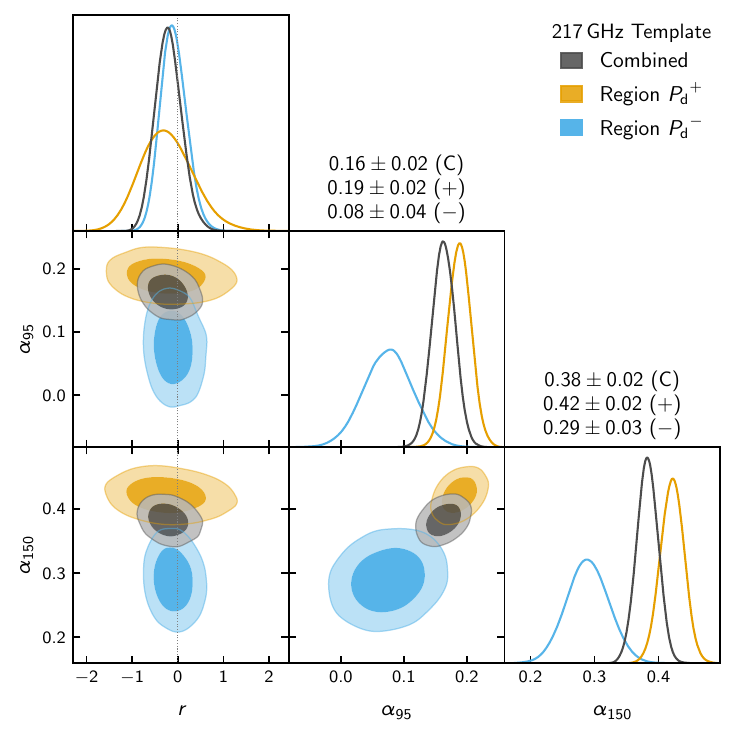}
  \caption{The XFaster likelihood contours for $r$, $\alpha_{95}$, and
    $\alpha_{150}$ with the combined \spider region in gray and \gdust and
    \ldust subregions in orange and light blue, respectively, computed
    independently with the 353\,GHz dust template (\textit{left}) and the
    217\,GHz dust template (\textit{right}).  Central values and $1\,\sigma$
    errors for each of the three regions are indicated in the titles for
    the $\alpha$ parameters.
    \label{fig:90x150a_spatial_var_triplot}}
\end{figure*}

Following the procedure described in
Section~\ref{sec:template_method}, we perform foreground template
subtraction \emph{independently} on each of the two subregions
defined in Section~\ref{sec:generating_subregions}.
The left panel of Figure~\ref{fig:90x150a_spatial_var_triplot} shows, for each subregion
and for the entire \spider field, the resulting 1-D and 2-D likelihoods for the
tensor-to-scalar ratio, $r$, and for the foreground-template fitting
parameters, $\alpha_{95}$ and $\alpha_{150}$, where the added indices
refer to the \spider frequencies to which we fit the \planck 353\,GHz foreground template.
Following the method in Section~4.3 of \citet{Gambrel_2021}, the results in this section are validated with a suite of 300 simulations with input
$r$, $\alpha_{95}$, and $\alpha_{150}$ fixed to the maximum likelihood values in Figure~\ref{fig:90x150a_spatial_var_triplot}.
Each simulated map is constructed from ensembles of CMB plus \spider noise, along with either the 217 or 353\,GHz \planck dust template.
Fitted templates add noise from \planck FFP10 realizations.
To produce robust estimates of the $\alpha$ parameters recovered from the \spider data, we correct the data likelihoods for a small ($\ll 1\,\sigma$) bias in the XFaster estimator, as measured from this validation suite.

With the \planck 353\,GHz template, we find the maximum-likelihood values for $r$ and $\alpha_{95}$ to be
consistent to better than $0.7\,\sigma$ and $2.0\,\sigma$, respectively,
between all three regions. However, the maximum-likelihood values for
$\alpha_{150}$ differ by $3.6\,\sigma$ between the two subregions, with a 
significantly higher value of $\alpha_{150}$ in \ldust than in \gdust.
In other words, the template-correlated dust signal in the more diffuse 
region, \ldust, is brighter than would be inferred by scaling the template 
based upon observations of the region with more polarized dust emission, \gdust (or of the combined region, 
where the fit is naturally driven by the brighter areas).
We would thus expect residual dust signal after subtraction of a template that 
did not account for such variability.

Note that these data are \emph{not} sufficient to establish any general causal relationship 
between variation in $\alpha_{150}$ and the intensity of the dust $P$ map. 
This would require comparisons
between a larger number of subregions, which is beyond the scope of
this work.

We are able to reject a number of compact structures as the
origin of this discrepancy. Two regions of particular interest are the
Orion-Eridanus Superbubble~\citep{Finkbeiner_2003,Soler_2018} and the
Magellanic Stream~\citep{Mathewson_1974}.
The former, which features
prominently in H$\alpha$ maps, occupies about 20\% of the full \Spider region, mostly located in \gdust.
Although the Magellanic Stream falls mostly outside
the \Spider region, some high-velocity clouds have been mapped on the
outskirts of its bottom-right corner~\citep{HI4PI_2016}. Even though
we do not necessarily expect the dust emission from these
regions to dominate the polarized emission in our maps, high levels of
polarization along the edge of the Orion-Eridanus Superbubble and
hints of variation in dust properties between the Galaxy and the
Magellanic Clouds~\citep{Galliano_2018} warrant caution. As documented
in Appendix~\ref{app:bubble+MCs}, we find the measured spatial
variation in $\alpha_{150}$ to be robust to masking aimed at probing
the impact of these structures.

We further explore the discrepancy in $\alpha$ between subregions by repeating this template subtraction analysis with an alternate 
dust template, constructed by substituting the 217\,GHz \Planck map for
the 353\,GHz map.
The 217\, GHz data also contain substantial polarized dust emission, as evidenced 
by the fact that they provide the second-largest contribution to the
SMICA dust reconstruction (Figure~\ref{fig:smica_all_weights}).
The right
panel of Figure~\ref{fig:90x150a_spatial_var_triplot} shows the 1-D
and 2-D likelihoods for $r$, $\alpha_{95}$, and $\alpha_{150}$ in this
configuration.  We again find $r$ to be consistent (within
$0.2\,\sigma$) between the two subregions, a moderate discrepancy in
$\alpha_{95}$ ($2.7\,\sigma$), and a larger discrepancy in
$\alpha_{150}$ ($3.5\,\sigma$).  However, while the latter discrepancy
has similar significance to that found with the 353\,GHz template, it
is in the opposite direction: $\alpha_{150}$ is found to
be \emph{lower} in \ldust than in \gdust. 

\subsection{Implications for a Modified-Blackbody Model}
\label{sec:spatial_variation_betad_Td}

\begin{figure}
  \centering
  \includegraphics[width=\columnwidth]{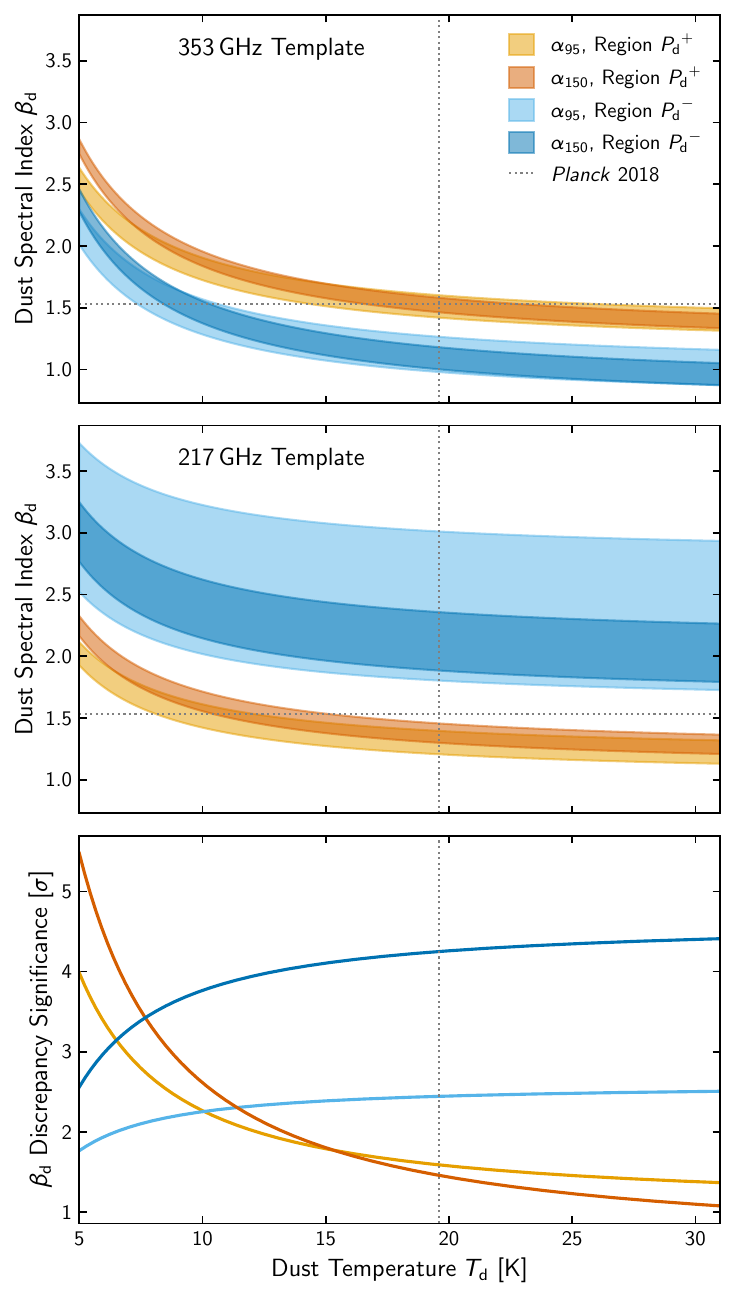}
  \caption{$1\,\sigma$ contours of possible $T_\mathrm{d}$ and $\beta_\mathrm{d}$
    for the $\alpha$ values found with the XFaster framework and presented in Figure~\ref{fig:90x150a_spatial_var_triplot}.
    (\textit{top}) Using the 353\,GHz dust template,
    the \gdust region is identified with a higher
    $\beta_\mathrm{d}$ or a higher $T_\mathrm{d}$, while the \ldust region is
    identified with a lower $\beta_\mathrm{d}$ or $T_\mathrm{d}$.
    (\textit{middle}) With the 217\,GHz template, the \gdust region is instead identified
    with a lower $\beta_\mathrm{d}$ or $T_\mathrm{d}$ than the \ldust region.
    (\textit{bottom}) 
    Mutual significance of the discrepancy in $\beta_\mathrm{d}$ between the two templates, as a function of assumed $T_\mathrm{d}$.
    The $\beta_\mathrm{d}$ values found in the \ldust region are discrepant at $>3\,\sigma$, while the significance of the discrepancy is $<2\,\sigma$ in the \gdust region.
    The \planck 2018 all-sky best fit parameters ($\beta_\mathrm{d} = 1.53$,
    $T_\mathrm{d} = 19.6$\,K) are also shown for reference (\textit{dotted lines}).
    \label{fig:beta_vs_tdust}}
\end{figure}

In the context of the simple MBB model of Equation~\ref{eq:MBB}, 
the template scaling parameter $\alpha$ is a function of the dust model parameters---the dust temperature, $T_\mathrm{d}$, and the spectral index of polarized emission, $\beta_\mathrm{d}$---as well as the frequencies of the maps used to build the dust template and the
frequency of the map to which the template is fitted.
Specifically,
\begin{equation}
  A_{\nu, \nu_0} =
  \frac{(e^{\,x} - 1)^2}{x^2 e^{\,x}}
  \frac{{x_0}^2 e^{\,x_0}}{(e^{\,x_0} - 1)^2}
  \left(\frac{\nu_0}{\nu}\right)^{\!2}
  \left(\frac{\nu^\mathrm{d}}{\nu_0^\mathrm{d}}\right)^{\!\beta_\mathrm{d}}
  \frac{B(\nu^\mathrm{d},T_\mathrm{d})}{B(\nu^\mathrm{d}_0,T_\mathrm{d})},
  \label{eq:dust_bigalpha}
\end{equation}
where the superscript ``d'' added to the frequency of a map indicates
that its effective band center is computed for a dust SED instead of
the CMB SED, as discussed in
Appendix~\ref{appendix:alpha_beta_conv}, and we have used the
shorthand $x\equiv h\nu/(k T_\mathrm{CMB})$. 

Using this relationship, we can translate the 1-D likelihoods for $\alpha_{95}$ and $\alpha_{150}$
shown in Figure~\ref{fig:90x150a_spatial_var_triplot} into contour plots in 
$(T_\mathrm{d},\beta_\mathrm{d})$.
The results are shown in
Figure~\ref{fig:beta_vs_tdust}.
These parameter constraints allow us to perform several internal comparisons 
within the MBB model context, which are discussed in the paragraphs below.

\paragraph{Comparing \Spider frequencies}
The $\beta_\mathrm{d}$ values inferred from the fitted $\alpha_{95}$ and 
$\alpha_{150}$ show excellent consistency (${<}1\,\sigma$) with one another for 
either choice of dust template, except perhaps in \gdust at the very lowest dust 
temperatures ($T_\mathrm{d} \lesssim 7\,\mathrm{K}$).
In other words, if we confine ourselves to a single \Planck-derived dust template, 
the \Spider data provide no significant evidence of deviation from an MBB SED in either subregion.

\paragraph{Comparing subregions}
The values of $\alpha_{150}$ obtained using the $\nu_0 =$~353\,GHz
dust template translate into \emph{lower} values of $\beta_\mathrm{d}$
in \ldust compared to \gdust for any value of
$T_\mathrm{d}$. Specifically, if we fix
$T_\mathrm{d}=19.6\,\mathrm{K}$ \citep{Planck_2018_XI} throughout, we
find $\beta_\mathrm{d} = 1.52 \pm 0.06$ in \gdust and $1.09 \pm 0.09$
in \ldust. Importantly, the value of $\beta_\mathrm{d}$ inferred
independently from $\alpha_{95}$ is consistent with that of
$\alpha_{150}$, with comparable significance.
Taken at face value, this $3.9\,\sigma$ inconsistency in $\beta_\mathrm{d}$ would indicate a
spatial variation in dust composition between the \gdust
and \ldust subregions.  

If we instead fix the dust spectral index to
be $\beta_\mathrm{d} = 1.53$ \citep{Planck_2018_XI} in both
subregions, we find a large difference in implied dust \emph{temperatures}:
$T_\mathrm{d}=19.1^{+3.7}_{-2.5}\,\mathrm{K}$ in \gdust, 
$T_\mathrm{d}=9.2^{+1.1}_{-0.8}\,\mathrm{K}$ in \ldust.  Although we certainly
expect \emph{some} lack of uniformity in dust temperature---stemming
from spatial variation in the dust's radiative environment---such a
large variation in the temperature of the diffuse interstellar medium
at high Galactic latitude may be unphysical given our current
understanding of dust physics \citep{draineli_2001, draine_2007}.

A similar discrepancy between subregions (though of \emph{opposite sign}) is visible with the 217\,GHz dust 
template, as shown in the middle panel of Figure~\ref{fig:beta_vs_tdust}.
Fixing 
$T_\mathrm{d}=19.6\,\mathrm{K}$ with
the $\alpha_{150}$ constraint results in $\beta_\mathrm{d} = 1.37 \pm 0.08$ in \gdust and 
$2.1^{+0.3}_{-0.2}$ in \ldust, a $3.2\,\sigma$ discrepancy. Alternatively, fixing
$\beta_\mathrm{d} = 1.53$ gives
$T_\mathrm{d}=12^{+3}_{-2}\,\mathrm{K}$ for \gdust, but 
no value of $T_d$ accommodates $\beta_\mathrm{d} = 1.53$ within \ldust.

\paragraph{Comparing dust templates within \ldust}
The values of $\beta_\mathrm{d}$
derived from $\alpha_{150}$ within the \ldust subregion using the $\nu_0 = 217$\,GHz dust template
are \emph{inconsistent} with those derived using the $\nu_0 = 353$\,GHz
template. The bottom panel of Figure~\ref{fig:beta_vs_tdust} shows the significance of this
discrepancy as a function of dust temperature: it exceeds $3\,\sigma$
($4\,\sigma$) for any $T_\mathrm{d}\gtrsim 5\,\mathrm{K}$
($T_\mathrm{d}\gtrsim 15\,\mathrm{K}$). 
Under the assumption that the
morphology of the polarized emission from interstellar dust is
identical in both templates, this result corresponds to a
statistically significant break in $\beta_\mathrm{d}$ for any physical
$T_\mathrm{d}$, with the index having a lower value at the higher
frequency.
Such an interpretation is complicated, however, by a comparison between subregions: the inferred values of $\beta_\mathrm{d}$ are \emph{lower} in 
\ldust than \gdust when using the 353\,GHz template, but \emph{higher} when using the 
217\,GHz template.
We can nonetheless infer a
statistically significant 
deviation from a single MBB spectrum within \ldust.

\paragraph{Discussion}

Taken together, the observations above point toward a more complex
picture of the polarized emission from interstellar dust in
the \spider region than that provided by a single-component MBB
model. 
To summarize: 
{\it i)} Unless $T_\mathrm{d}$ exhibits large variations at
high Galactic latitudes, the spatial variation in $\alpha$
between \gdust and \ldust may be evidence of \emph{multiple dust
populations} (\emph{i.e.}, differing compositions, and thus emissivities);
{\it ii)} The inconsistency in the value of $\alpha$ derived in \ldust with one template 
and that found in the same subregion with the other template requires either a 
\emph{deviation from an MBB spectrum} or a 
\emph{discrepancy in the morphology of polarized dust emission} between the two templates, or both.
Such a discrepancy in morphologies could arise from a spatially varying dust SED, \emph{e.g.}, due to the presence of 
multiple dust populations.
The possible presence of multiple dust populations within \ldust is perhaps not surprising, given the spatial variation in $\alpha$
between \ldust and \gdust detected with the  $\nu_0=353$\,GHz template.

Evaluating this rather complex picture will require more data. 
Deep observations between 150 and 353\,GHz (such as the \spider 280\,GHz maps) will be 
particularly valuable 
in distinguishing between frequency-dependent dust morphology and complexity in the dust SED.
If deviations from an MBB model are borne out,
this may provide new insights into the
properties of the interstellar medium, and associated modeling
uncertainties would need to be taken into account in upcoming searches
for cosmological $B$-modes.

%% file: testing_dust_modeling.tex
The SMICA framework offers a high degree of flexibility in modeling the
characteristics of dust.  The SMICA pipeline implemented for the analysis in
\citet{bmode_paper} incorporates a MBB model for the dust foreground component
with the following characteristics:
\begin{itemize}
\setlength\itemsep{0em}
\item A single $\beta_\mathrm{d}$ parameter that is common across all multipoles and polarizations,
\item A dust amplitude $A_\mathrm{d}$ that is free to vary across multipoles and polarizations,
\item A $\delta$-function prior of $T_\mathrm{d} = 19.6\,\textrm{K}$ to fix the dust temperature at the \planck all-sky value, and
\item A Gaussian prior on $\beta_\mathrm{d}$ centered at the \planck all-sky value
  $\beta_\mathrm{d} = 1.53$, with a generous width of $\sigma_{\beta_\mathrm{d}} = 1.5$.
\end{itemize}
In this section, we quantify how these foreground modeling choices impact the
best-fit CMB and foreground parameters recovered with the combined \spider and
\planck data.  Throughout this section, we leave the dust temperature fixed,
due to a lack of sufficient constraining power and frequency coverage
to simultaneously estimate both $\beta_\mathrm{d}$ and $T_\mathrm{d}$,
while varying the constraints on $A_\mathrm{d}$ and/or
$\beta_\mathrm{d}$.  Here, unlike the discussion in
Section~\ref{sec:spatial_variation_betad_Td}, we apply the analysis to the full \Spider region.  The results in this section are therefore driven by the higher signal-to-noise component, corresponding approximately to the \gdust contours of Figure~\ref{fig:beta_vs_tdust}.

\subsection{SED Modeling of Polarized Dust Emission}

\begin{table}[t]
  \caption{The best-fit dust amplitude $A_\mathrm{d}$ for each of the input
    frequency maps relative to that of the \Planck 353\,GHz map. The values
    tabulated here are also plotted in Figure~\ref{fig:dust_amp_v_betad}. Best-fit
    values are computed as the median of the posterior distribution, with $1\,\sigma$
    errors determined from the 68\% interval on either side of the median.
    \label{table:dust_amp_v_betad}
  }
  \hspace{-15mm}\begin{tabular}{lc|r@{ }c@{ }lr@{ }c@{ }l}
    \toprule
    & & \multicolumn{6}{c}{Dust Amplitude} \\
    Instrument & Band & \multicolumn{3}{c}{EE} & \multicolumn{3}{c}{BB} \\
    \midrule
    \spider & 95\,GHz & $0.019$ & $\pm$ & $0.005$ & $0.016$ & $_{-}^{+}$ & $_{0.004}^{0.005}$ \\
     & 150\,GHz & $0.047$ & $_{-}^{+}$ & $_{0.004}^{0.005}$ & $0.047$ & $\pm$ & $0.003$ \\
     \midrule
    \planck & 100\,GHz & $0.018$ & $_{-}^{+}$ & $_{0.006}^{0.005}$ & $0.028$ & $\pm$ & $0.007$ \\
    & 143\,GHz & $0.043$ & $\pm$ & $0.008$ & $0.039$ & $\pm$ & $0.010$ \\
    & 217\,GHz & $0.139$ & $\pm$ & $0.013$ & $0.123$ & $\pm$ & $0.014$ \\
    \bottomrule
  \end{tabular}
\end{table}

\begin{figure}
  \centering
  \includegraphics[width=\columnwidth]{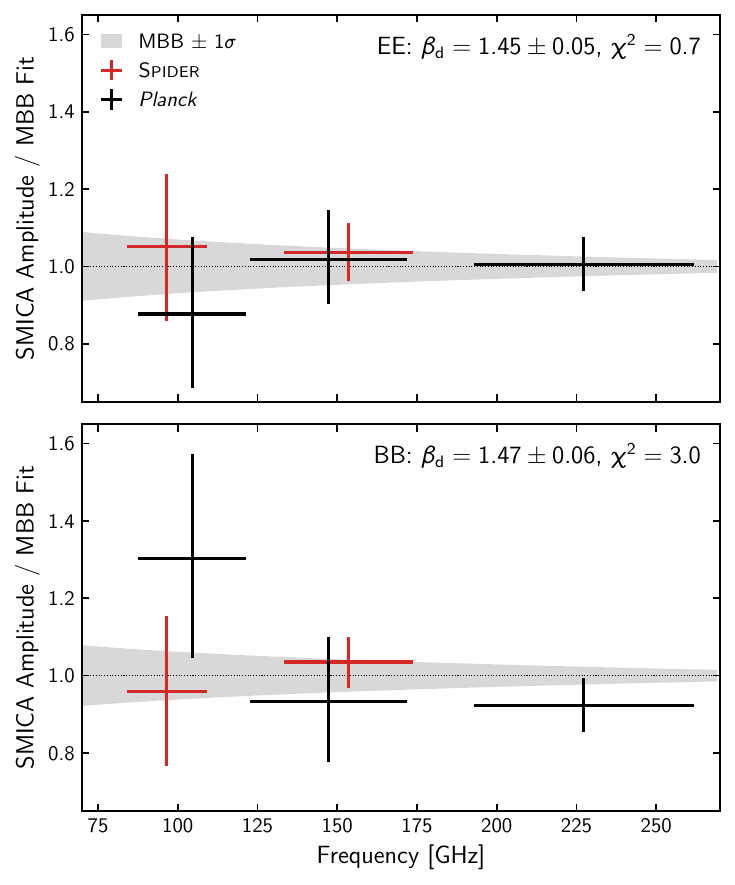}
  \caption{Testing the fidelity of the MBB dust model in the \spider
    high-latitude region for $E$- (\textit{top}) and $B$- (\textit{bottom})
    modes. All data are plotted relative to the best-fit MBB model for that
    polarization component (gray band indicates the $1\,\sigma$ fit envelope),
    and account for the detector bandpass at each frequency.  The horizontal error bars indicate the
    effective detector bandwidth over which the amplitudes are computed. The
    SMICA best-fit dust amplitudes ($+$) are consistent with the best-fit MBB
    model.  The \spider data (\textit{red}) have smaller error bars on $A_\mathrm{d}$,
    and therefore more constraining power, than the \planck data
    (\textit{black}) in nearby bands.
    \label{fig:dust_amp_v_betad}}
\end{figure}

We begin by testing the SED model of dust emission by fitting an independent
amplitude $A_\mathrm{d}$ at each frequency, operating under the assumption that this
amplitude remains consistent across multipoles.  As discussed in
\cite{bmode_paper}, dust and noise exhibit similar scaling behaviors with frequency,
making it challenging for the SMICA 
algorithm to distinguish between the two components.
By relaxing the prior constraint on $\beta_\mathrm{d}$ and fitting for the
dust amplitudes directly, we are able to circumvent this noise/dust scaling
degeneracy.  The resulting amplitudes can then be fit to a MBB model to recover
the best-fit $\beta_\mathrm{d}$ value.

The dust amplitudes recovered with this algorithm are listed in
Table~\ref{table:dust_amp_v_betad} for each of the \spider and \planck frequency
bands, and the resulting MBB model fit is illustrated in
Figure~\ref{fig:dust_amp_v_betad}.  Fitting a spectral index to
each set of six frequency maps results in a best-fit value of $\beta_\mathrm{d}$ of $1.45\pm0.05$
($1.47\pm0.06$) for the $EE$ ($BB$) components, with $\chi^2$ values of 0.7 (3.0),
respectively.\footnote{For comparison, we note that the template fit in the combined region from Section~\ref{sec:spatial_variation_betad_Td} corresponds to a value $\beta_\mathrm{d}=1.50\pm 0.05$ for the combined $EE$ \& $BB$ spectra.} With approximately four degrees of freedom, these $\chi^2$ values
indicate that a single-component MBB model is an adequate description of the
\spider + \planck dataset, thus justifying the use of the MBB model for
constraining foreground power in the CMB $B$-mode analysis.
Additionally, Figure~\ref{fig:dust_amp_v_betad} illustrates the improved
constraining power of the \spider data at 95~and~150\,GHz in this region of the sky
relative to that of the nearby \planck bands at 100~and~143\,GHz.

\subsection{Angular Scale Dependence of Dust Spectral Index}\label{subsec:scale_dep}

\begin{figure}
  \centering
  \includegraphics[width=\columnwidth]{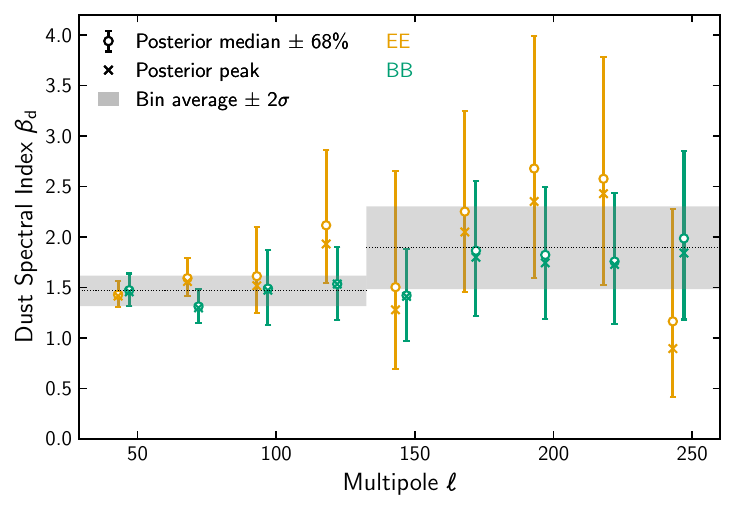}
  \caption{An analysis of the variation of the dust spectral index over angular
    scale using the SMICA pipeline. Each point covers a binned multipole range
    of $\Delta \ell = 25$, with $E$ and $B$ points offset in $\ell$ for
    clarity. The distributions of the best fit $\beta_\mathrm{d}$ in each bin are not
    Gaussian; the median ($\circ$) and peak ($\times$) of each posterior are indicated. The error bars correspond to 68\% intervals on either side of
    the median, equivalent to 1\,$\sigma$ for a non-Gaussian distribution.
    The best-fit $\beta_\mathrm{d}$ values agree to within a mutual $2\,\sigma$ between the larger and smaller angular scales.
    The $E$ and $B$ $\beta_\mathrm{d}$ values are
    statistically consistent with each other at all scales.
    \label{fig:betad_vs_ell}}
\end{figure}

Next we test the assumption of a scale-independent SED by fitting the MBB
model independently for each polarization as a function of $\ell$.  The results
of this fit for $\beta_\mathrm{d}$ are shown in Figure~\ref{fig:betad_vs_ell}.
The best-fit spectral index values agree to within $\sim 0.5\,\sigma$ between the
$E$ and $B$ polarization components across all angular scales, providing no
evidence for including a polarization-dependence in the model.
Comparing the best-fit $\beta_\mathrm{d}$ values averaged over larger ($\ell < 140$) and smaller
($\ell > 140$) angular scales shows agreement to within $2\,\sigma$.

When analyzed over the full \spider region, the data do not reveal any strong evidence of variations in spectral
index with polarization mode or angular scale, further justifying the adoption of the
simple MBB dust model.

\subsection{Angular Scale Dependence of Dust Amplitude}

\begin{figure}
  \centering
  \includegraphics[width=\columnwidth]{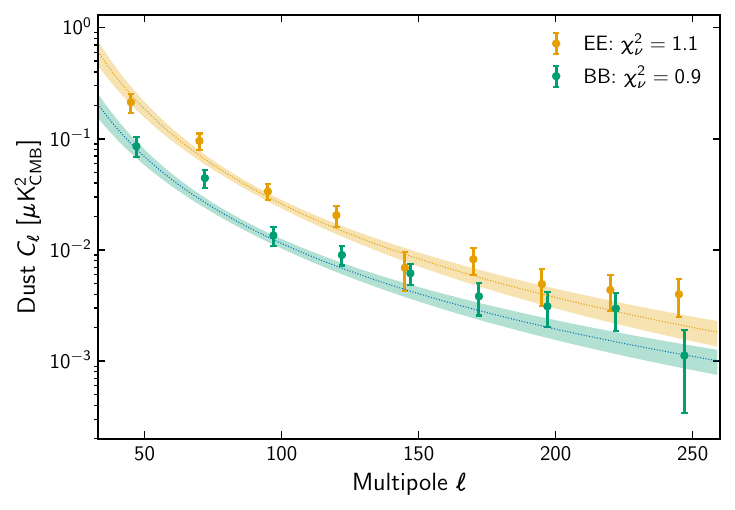}
  \caption{A power-law model fit to the SMICA-derived dust amplitudes as a
    function of angular scale. The best-fit model, shown as a band
    indicating the $1\,\sigma$ error on the fit, is binned identically to the
    SMICA-derived dust bandpowers in order to compute the $\chi^2$ of the fit.
    The reported $\chi^2_\nu$ values are computed using the full covariance
    of both the power-law model and the SMICA bandpowers, and assume seven
    degrees of freedom.  The best-fit model parameters are shown in
    Figure~\ref{fig:dust_powerlaw_cov}.
     \label{fig:dust_powerlaw_fit}}
\end{figure}

\begin{figure}
  \centering
  \includegraphics[width=\columnwidth]{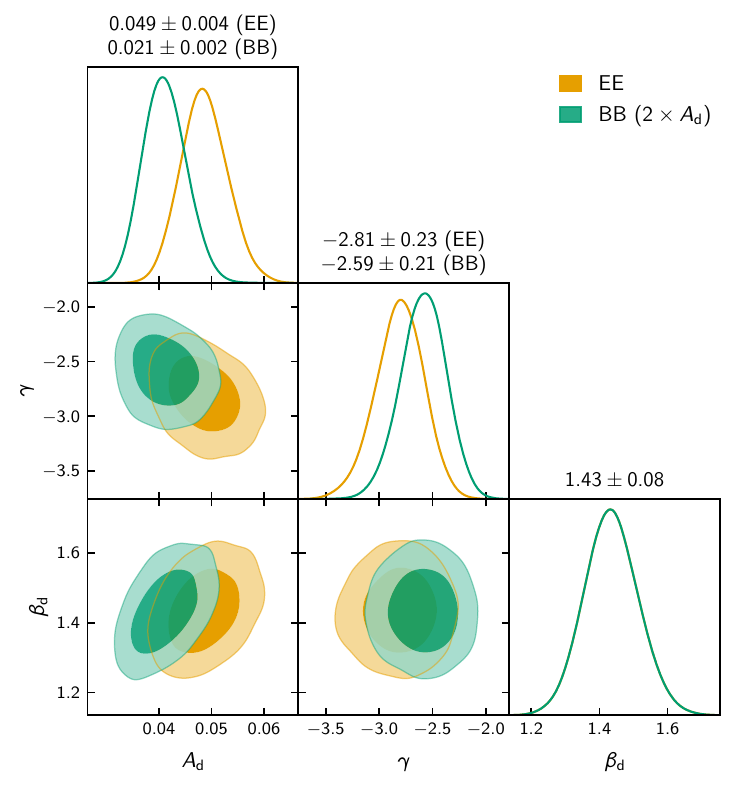}
  \caption{The covariance of the fitted dust parameters for the model shown in
    Figure~\ref{fig:dust_powerlaw_fit}.  Central values with $1\,\sigma$ errors
    are indicated in the titles for each of the three parameters. The best-fit
    $A_\mathrm{d}$ and $\sigma_{A_\mathrm{d}}$ for the $BB$ component is scaled
    by a factor of 2 for better visual comparison to its $EE$ counterpart, with
    its true central value indicated in the column title.
    \label{fig:dust_powerlaw_cov}}
\end{figure}

Dust emission is often (\emph{e.g.}, \citet{Planck_2018_XI,BK18_2021}) modeled as a power law in $C_\ell$:
\begin{equation}
  C_\ell = A_\mathrm{d}(\nu) \left( \frac{\ell}{80} \right)^{\gamma}.
\end{equation}
We can fit this model to the SMICA-derived amplitudes as a function of $\ell$-bin
independently for the $E$ and $B$ polarization components. We assume a
single-component MBB scaling of the amplitude with frequency, common to both
polarizations. The results of this fit are shown in
Figures~\ref{fig:dust_powerlaw_fit} and~\ref{fig:dust_powerlaw_cov}, with reduced
$\chi_\nu^2 = 1.1 (0.9)$ for the $E$ ($B$) polarization components, assuming $\nu = 7$.
This analysis finds a dust spectrum that falls steeply, $\gamma^{EE} = -2.81\pm0.23$, with
no significant difference in best-fit values between the two polarizations.  We note that \Planck has reported values of $-2.42$ and $-2.54 \pm 0.02$ in the largest area of sky for $EE$ and $BB$, respectively.  In the most restrictive mask, \Planck finds more shallow indices of $-2.28 \pm 0.08$ and $-2.16 \pm 0.11$ for $EE$ and $BB$ \citep{Planck_2018_XI}.

The \Spider dataset also provides an additional observational constraint on the asymmetry in
dust power between $EE$ and $BB$ polarization modes, visible in Figure~\ref{fig:dust_powerlaw_cov}.
This analysis finds an $EE/BB$ dust power ratio of $2.1 \pm 0.3$ at large scales ($\ell < 140$).
This result aligns with prior findings from \Planck
\citep{planck_XXX_2014}. The origin of this dust power asymmetry is not
conclusively understood.  Some studies investigate the role of
magnetohydrodynamic (MHD) turbulence in the interstellar medium
(ISM) \citep{caldwell_2017, kandel_2017}, while others suggest that
this power asymmetry can be a probe of the dust filament aspect
ratio \citep{huffenberger_2020} and magnetic field alignment \citep{clark2021}.

\subsection{Impact on Cosmological $B$-mode Measurements}
\label{sec:impact_on_sigmar}

At the level of sensitivity of contemporary observational programs,
the contribution of modeling uncertainty to limits on the cosmological
CMB $B$-mode signal is non negligible even in those regions of the sky
with low levels of Galactic emission.  In this section we attempt to
quantify the impact of these complexities on the uncertainty in the
tensor-to-scalar ratio, $\sigma_r$, within the \spider region.

Following the analysis in \citet{bmode_paper}, for each dust model we
can compute the $\sigma_r$ derived from an ensemble of sky and noise
simulations constructed to match the model under consideration.  In
the configuration where there is no SED modeling and SMICA fits a dust
amplitude per frequency, we find $\sigma_r = 0.130$.  If instead a
spectral index $\beta_\mathrm{d}$ is fit per multipole, allowing for
dust behavior to change with scale, we find $\sigma_r = 0.128$.
Constraining the spectral index to a single value over all scales, as
in \cite{bmode_paper}, we find $\sigma_r = 0.116$.  Finally, the
assumption that dust follows a power-law in multipole space \emph{does
not} improve $\sigma_r$ further.  We find instead that this assumption
re-distributes the uncertainties: the uncertainty near the pivot scale
($\ell \sim 80$) is reduced, while that at larger scales increases.
Thus, we find that the complexities in foreground modeling described in this section have a ${\sim}10\%$ impact on  $\sigma_r$ in the case of \Spider.

\subsection{Discussion}

When applied to the full area of sky observed by the \spider experiment, the variations of the nominal SMICA algorithm described
above, each employing different assumptions about foreground modeling, all point
to a consistent description of polarized Galactic dust emission. These data are consistent with dust characterized by a single-component, modified black-body model across the \spider and \planck frequencies and for both polarization parity modes.  The dependence on angular scale is consistent with an
approximate power-law dependence on multipole, $\ell$.  The combined \spider and \Planck data over this restricted region prefer a dust emissivity spectral index $\beta_\mathrm{d} = 1.45\pm0.07$, somewhat lower than the all-sky \planck 2018
value \citep{Planck_2018_XI} of $\beta_\mathrm{d} = 1.53 \pm 0.02$ with $T_\mathrm{d} = 19.6K$.
The foreground complexities considered in this section are found to contribute to the uncertainty on the tensor-to-scalar ratio at the ${\sim}10\%$ level;
we emphasize, however, that the analysis in this section does not explore the effects of the possible spatial variation discussed in Section~\ref{sec:spatial_variation}, for which we lack a single compelling model.

%% file: dust_angle.tex
In addition to the search for spatial variation in Section~\ref{sec:spatial_variation}, we use \Spider data to perform a search for variation in the polarization angle of dust emission with frequency.
Such a rotation with frequency may indicate the presence of multiple dust clouds along a line of sight with different SEDs and magnetic alignments.
For example, \citep{pelgrims_2021} used HI line emission maps to identify regions of the sky likely to contain magnetically-misaligned clouds; they compute the difference in polarized dust angle for these different populations to demonstrate that LOS frequency decorrelation is detectable within the Planck dataset, largely in the Northern Galactic cap. An analysis by the BICEP team \citep{bicep_XVI} showed no evidence of dust decorrelation within their observation region.

We search for LOS decorrelation by evaluating the consistency between a set of maps of estimated dust polarization angle, each derived from a single \spider or \planck map. 
We first construct a set of dust templates as in Section~\ref{sec:template_method}. We focus on the \planck 353\, GHz, \planck 217\, GHz, and \spider 150\,GHz maps, as these have the highest sensitivity to polarized dust. 
For this analysis we opted to subtract CMB signal using the SMICA-derived CMB template rather than the \Planck~100\,GHz map, due to its lower overall noise level. %
From each template, we compute a map of polarized dust angle using the unbiased estimator~\citep{plaszczynski_polest}
\begin{equation}
  \psi = \frac{1}{2} \arctan(U,Q).
  \label{eq:dust_angle}
\end{equation}
We then construct a likelihood estimator to evaluate the consistency of the measured dust angle $\psi$ at different frequencies.
As discussed further below, we employ a Gaussian mixture model (GMM) \citep{hogg_likelihood} approach that models the data as drawn from a mixture of two populations (consistent and inconsistent across frequencies) and assigns each pixel a probability of belonging to each.
Relative to a standard $\chi^2$ consistency analysis, this probabilistic assessment can identify regions of substantial disagreement without requiring an arbitrary choice for an outlier threshold. The self-tuning nature of a GMM is particularly appealing given the low per-pixel signal-to-noise on dust angle.

In order to construct our likelihood, we approximate the uncertainty on $\psi$ as Gaussian.
We estimate this uncertainty with standard error propagation:
\begin{multline}
  \sigma_\psi^2 =  \frac{1}{4} \frac{Q^2 \sigma_U^2 + U^2 \sigma_Q^2}{(Q^2 + U^2)^2} \\ + \frac{Q^2U^2}{2(Q^2 + U^2)^4} \left[ \sigma_Q^4 + \sigma_U^4 + \frac{1}{2}\frac{(Q^2 - U^2)^2}{Q^2U^2}\sigma_Q^2 \sigma_U^2 \right]
\end{multline}
where the two grouped terms represent the first- and second-order errors respectively.
We expect these uncertainties to be underestimated when the signal-to-noise is low ($\sigma_U^2 + \sigma_Q^2 > U^2 + Q^2$). However, the Gaussian approximation breaks down as higher-order terms become more significant. To address this, we apply a mask to reject map pixels for which the second-order errors on polarized dust angle exceed the first. This criterion rejects 20-30\% of the sky area covered by the \spider instrument.
This thresholding serves two main purposes: (1) to exclude pixels whose uncertainties are challenging to model accurately, and (2) to remove pixels with especially large uncertainties, for which the width of the distribution would wrap around the range of the arctangent function, yielding non-Gaussian behavior. While the statistical properties of $\psi$ are complex and cannot be strictly Gaussian, the Gaussian distribution serves as a reasonable approximation after this thresholding.

\begin{figure}
  \centering
  \includegraphics[width=\columnwidth]{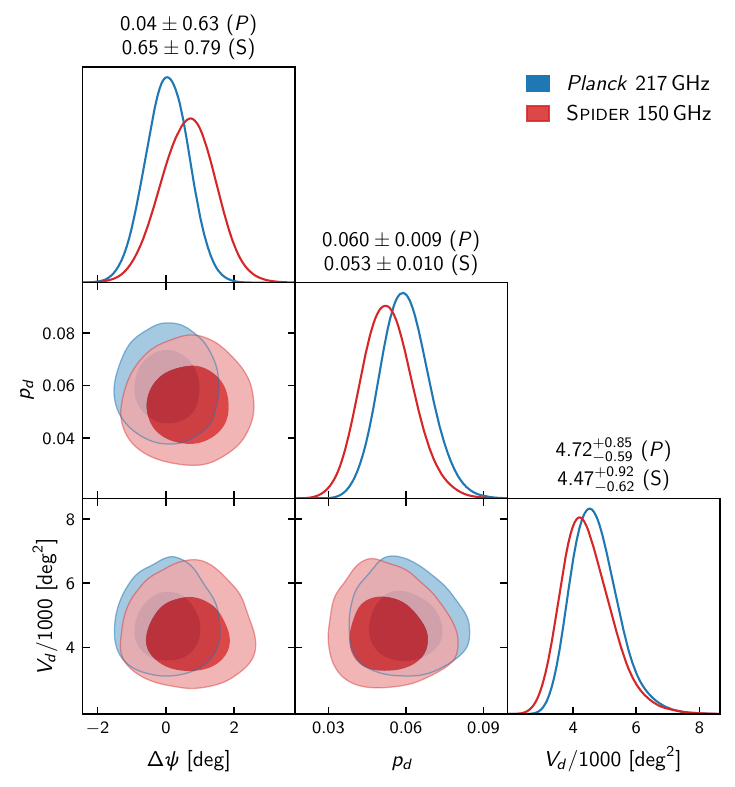}
  \caption{Best-fit model parameters for the dust angle decorrelation likelihood
  defined in Equation~\ref{eq:like_decorrelate}: global dust angle offset
  $\Delta\psi$ in degrees, population fraction $p_d$ of pixels with measurable angle differences,
  and additional variance $V_d$ in square degrees introduced by these outlier pixels.
  We find that $\Delta\psi$ is statistically consistent with zero, indicating
  that the dust angle appears globally consistent between the \planck 353\,GHz
  frequency map and lower frequencies \planck 217\,GHz (\textit{blue})
  and \spider 150\,GHz (\textit{red}).  The likelihoods are marginalized over $\{q_i\}$,
  and central values with $1\,\sigma$ errors are indicated in the titles.
  \label{fig:dust_angle_decorr_corner}}
\end{figure}

\begin{figure}
  \centering
  \includegraphics[width=\columnwidth]{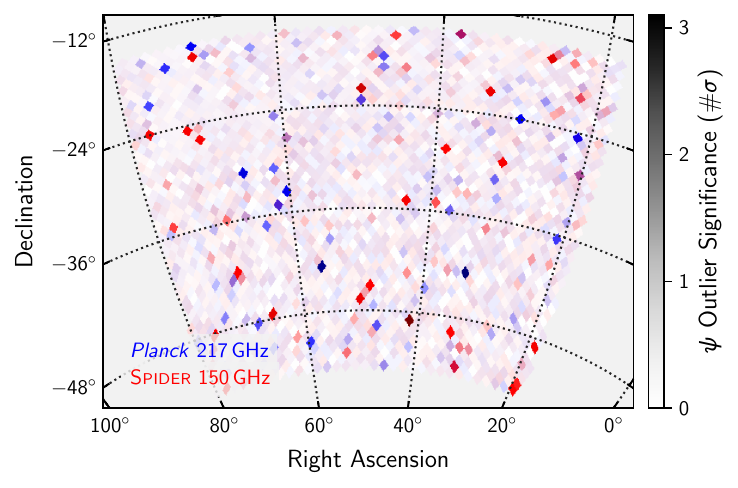}
  \caption{The significance in $\sigma$ of the probability $p_i$ that the
  measured dust angle in each map pixel $i$ is inconsistent between the \planck
  353\,GHz map and either the \planck 217\,GHz map (\textit{blue}) or
  the \spider 150\,GHz map (\textit{red}). The significance is encoded such that
  bright purple pixels are likely in the outlier population in both frequency
  pair analyses.  The lack of both astrophysical structure and overlap between
  the two populations argues against local decorrelation in the \Spider
  observing region.
  \label{fig:dust_angle_decorr_map}}
\end{figure}

From a given pair of dust maps $X$ and $Y$ we calculate a pair of dust angles ($\psi^X_i,\psi^Y_i$) for each map pixel $i$.
To this we fit a GMM in which each pixel can belong to one of two populations: a base population, for which the dust angle is \emph{consistent} between our pair of maps up to some constant angular offset $\Delta \psi$; and a second population of outlier pixels, for which the polarization angle is \emph{inconsistent} between the two frequencies. 
The first population probes a possible systematic offset in polarization angle between maps at different frequencies; this is expected to be consistent with zero, and unlikely to arise from dust decorrelation given the relatively large sky area covered by \spider.
The second population is our probe of decorrelated dust, in the form of isolated, dense cloud structures with arbitrary polarization angles relative to our line of sight.

Our likelihood takes the form
\begin{multline}
  \mathcal{L}(\{q_i\}, \Delta \psi, p_d, V_d) = \\
  \prod_i \left[ \frac{1}{\sqrt{2 \pi \Sigma_i}} \exp \left( -\frac{\Delta_i^2}{2 \Sigma_i} \right) \right]^{1 -q_i} \times \\
  \left[\frac{1}{ \sqrt{2 \pi (\Sigma_i + V_d)}} \exp \left( -\frac{\left(\psi^X_i\right)^2 + \left(\psi^Y_i\right)^2}{2 (\Sigma_i + V_d)} \right) \right]^{q_i},
  \label{eq:like_decorrelate}
\end{multline}
where
\begin{align}
  \Delta_i &= \min_{\delta_\pi} \left \{ \frac{1}{\sqrt{2}} \left( (\psi^Y_i + \delta_\pi) - (\psi^X_i + \Delta \psi) \right)  \right \}\label{eq:ortho_distance} \\
  \Sigma_i &= \frac{1}{2} \left(\sigma_X^2 +\sigma_Y^2\right) \\
  q_i & \sim \text{BetaBinomial}\left(\alpha = p_d, \beta = 1 - p_d\right)
\end{align}
The boolean $q_i$ assigns each pixel $i$ to one of the two populations.
The generative model of the base population ($q_i=0$) is taken to be a two-dimensional Gaussian, with an orthogonal distance $\Delta_i$ between the two map angles for each pixel (Equation~\ref{eq:ortho_distance}) and variance $\Sigma_i$ (the projected variance of the uncertainty of the pixel onto a line of slope of unity). Note that, because the domain of angles lies in the range $[-\pi/2, \pi/2]$, differences in $\psi$ by any multiple of $\pi$ are inconsequential.
We account for this by minimizing over a nuisance parameter $\delta_\pi \in \{ -\pi, 0, \pi \}$ in Equation~\ref{eq:ortho_distance}.

The second population represents outlier, or rejected, pixels, which show local differences in angle between the maps. We do not \emph{a priori} have an expectation for the statistics of this population. We choose to model it as a two-dimensional Gaussian with zero mean and an additional variance $V_d$ (in excess of $\Sigma_i$).

Finally, we treat the pixel assignments $q_i$ probabilistically, such that each pixel has a probability $p_i$ of being assigned to the base population.
The probability $p_i$ is itself drawn from a beta distribution with $\alpha  = p_d, \beta = 1- p_d$, such that the expectation value $E(p_i) = p_d$. In this way, $p_d$ can be thought of as a global parameter for the fraction of outlier pixels in the total population. All together, our model has $N_\mathrm{pix}+3$ variables: $N_\mathrm{pix}$ population assignments $q_i$, a systematic offset parameter $\Delta \psi$, and two parameters describing the variance ($V_d$) and population fraction ($p_d$) of the outliers.

We conduct two iterations of this analysis using pairs of maps with the highest signal-to-noise ratios for dust: one comparing the \Planck 353\,GHz and \Spider 150\,GHz maps, the other comparing \Planck 353\,GHz to \Planck 217\,GHz.
Figure~\ref{fig:dust_angle_decorr_corner} shows the resulting uncertainties and covariances between the fit parameters $(\Delta \psi, p_d, V_d)$.
Both comparisons find angular offsets $\Delta \psi$ consistent with zero to within $1\,\sigma$, with $p$-values of 0.21 (0.47) for $\Delta \psi_{150}$ ($\Delta \psi_{217}$).
This suggests that the mean polarized dust angle at 353\,GHz is consistent with those at lower frequencies.

Only about 5\% of the pixels included in our analysis region are rejected and assigned to the outlier population. These pixels display discrepancies in dust angle between 353\,GHz and the corresponding lower frequency.
While these outliers could signify spatially-dependent decorrelation, a map of the locations of these pixels (Figure~\ref{fig:dust_angle_decorr_map}) reveals two key observations: (1) the regions lack the connectivity we might expect from a real astrophysical population, and (2) the outliers identified by the two analyses share little overlap. Indeed, only about 0.2\% of pixels are identified as deviant in both analyses. This argues against interpreting these pixels as a detection of LOS decorrelation.

We can further perform a hypothesis test to estimate the probability of this analysis yielding 5\% decorrelated pixels under the null hypothesis that there is no decorrelated dust at any frequency for the entire observed region.
We construct a dust map from 353-100\,GHz difference maps, then scale this to both 150\,GHz and 217\,GHz using a standard modified black-body dust model.
To this we add realistic noise by drawing from an ensemble of simulations, specifically  \spider noise simulations at 150\,GHz and \planck FFP10 realizations at 217\,GHz and 353\,GHz. From these simulated dust + noise maps, we compute the dust angles and perform an identical analysis.
Comparing our measurement of the fraction of outlier pixels, $p_d$, to this ensemble of simulations, we obtain $p$-values of 0.18 and 0.10, respectively, for the 150\,GHz and 217\,GHz analysis.
We thus conclude that the inferred fraction of decorrelated pixels can be readily attributed to noisy measurements of Stokes $Q$ and $U$.
Consequently, our analysis finds no evidence for LOS decorrelation in the \spider observing region. The dust angles at 353\,GHz largely agree with those at the lower frequencies, and those pixels that show outsized deviations are plausibly caused by noise.

%% file: comp_to_pysm.tex
The SMICA dust maps derived from the combination of \Spider and
\Planck data (shown in Figure~\ref{fig:smica_dust_maps}) can also be
used to test full-sky dust models. The Python Sky Model, commonly
known as PySM, is a Python package designed for generating full-sky
simulations of Galactic microwave foregrounds \citep{Thorne_2017}.  It
is often used by the CMB community to test component separation
methods and perform sensitivity forecasts. The current release, PySM3,
contains thirteen distinct dust models that vary the assumptions and data
processing of the underlying \planck data, allowing the user to
generate full-sky dust emission maps at any frequency bandpass of
interest.

To test the agreement between each of these models and our SMICA dust
polarization maps, we first use PySM to generate maps in the \spider
observing region using the measured 150\,GHz frequency bandpass.
We then compute binned $EE$ and $BB$ spectra for each PySM model
using PolSPICE \citep{polspice} and compare these to analogous spectra generated
from the \Spider-\Planck SMICA dust maps described in
Section~\ref{subsec:SMICA}. %

\begin{figure}
  \centering

  \includegraphics[width=\columnwidth]{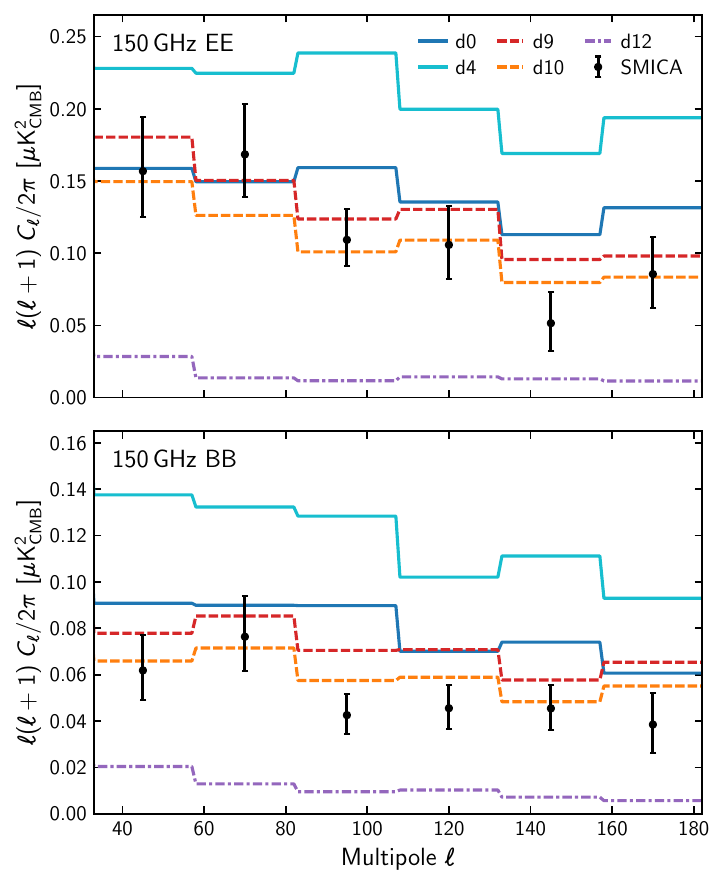}
  \caption{Comparison of the $EE$ (\textit{top}) and $BB$
    (\textit{bottom}) spectra from several PySM3 dust models
    (\textit{colored lines}) to SMICA-derived dust bandpowers at
    150\,GHz (\textit{black points}) in the \spider region.  Models
    based on the same underlying inputs and approach to SED modeling are shown
    with the same line style.
    \label{fig:pysm_comparison}}
\end{figure}

Figure~\ref{fig:pysm_comparison} shows these spectra for a representative
subset of dust models, which were chosen to span the range of
predicted dust amplitudes in the \spider observing region and
highlight several that have been used in recent experiment forecasts
(\textit{e.g.} \citet{SO_science_book, S4science}).  Briefly, each of
these are described as follows:
 \begin{itemize}
  \setlength\itemsep{0em}
 \item {\bf d0}: A single-component MBB with fixed $(\beta_\mathrm{d} = 1.54,\:T_\mathrm{d} =20\:K)$ using the \planck 2015 353\,GHz polarization map \citep{planck2015_X} as a template.
 \item {\bf d4}: A generalization of the MBB model with spatially varying temperature and spectral index based on the \planck 2015 results, scaled using the two-component model from \citet{finkbeiner1999extrapolation}.
 \item {\bf d9}: A single-component MBB model with fixed $(\beta_\mathrm{d} = 1.48,\: T_\mathrm{d} = 19.6\:K)$ based on the \planck 2018 results \citep{Planck_2018_XI}, which uses GNILC component-separated maps as a template \citep{planck2018_XII} with additional small-scale fluctuations.
 \item {\bf d10}: A variant of \textbf{d9} that also includes small-scale fluctuations in $\beta_\mathrm{d}$ and $T_\mathrm{d}$.
 \item {\bf d12}: A 3-D model of polarized dust emission with six layers as further described in \citet{dustMKD}.
 \end{itemize}
More information about each of these models can be found in
 the PySM documentation.\footnote{
   \url{https://pysm3.readthedocs.io/en/latest/models.html\#dust}}
The remaining PySM3 models are compared in Appendix~\ref{app:more_pysm} and Figure~\ref{fig:more_pysm_comparison}.

From these figures we can make some qualitative observations about how these models behave
in this region of the sky.
For instance, the similarity in predicted power between
models {\bf d0} and {\bf d9} suggests that differences in
the model parameters and templates between the \planck 2015 and 2018 releases do not
greatly affect the degree-scale dust predictions in this region of sky.
The scale of variation between {\bf d10} and {\bf d11} reflects
the impact of stochastic modeling of the small scale features.
On the other
hand, despite their common underlying \Planck data, several of these models
predict substantially different dust levels from the others in this region.
In particular, {\bf d4} predicts significantly more
degree-scale polarized dust emission than the other models while {\bf d12}
predicts significantly less, though a detailed exploration of the
origin of these differences is beyond the scope of this
work.

Comparing the PySM spectra with those from our SMICA analysis
is complicated by the fact that both SMICA and the various PySM models draw heavily upon the same underlying \Planck maps and noise ensembles.
In our own SMICA analysis, for example, the \Planck 353\,GHz map dominates the relative weights used to construct the dust component and thus drives the
morphology of the dust map and the {\em shapes} of the dust
power spectra.
The \emph{amplitudes} of those power spectra at
150\,GHz, however, are driven primarily by the \Spider data.
Since different PySM models employ the \Planck maps in different ways and combinations,
we expect complex covariances among all of these spectra that are difficult to quantify reliably.
We can nonetheless qualitatively observe closer correspondence of our on-sky results with some
models (\emph{e.g.}, {\bf d0}, {\bf d9}, {\bf d10}) than others (\emph{e.g.}, {\bf d4}, {\bf d12}).

%% file: conclusion.tex
This paper addresses the impact of choices made in component
separation for the \Spider cosmological $B$-mode analysis, and explores
the complexity of the polarized dust emission in the \Spider region.
A quantitative comparison of a template-based and
harmonic-independent-component analysis shows that, at the level
of the statistical noise in the \spider dataset, the modeling errors resulting from
the approximations made in each result in a small, but non-negligible
change in $\sigma_r$.

Template-based and harmonic domain methods make
entirely different assumptions about the behavior of
foregrounds. The value of employing multiple component
separation pipelines lies in the capacity to explore the properties of
foregrounds while evaluating the sensitivity of the result to
different analysis choices.  Future $B$-mode observational programs
will need to carefully consider the impact of these modeling choices
on the likelihood of the tensor-to-scalar ratio.

In Section~\ref{sec:spatial_variation}, we analyze the spectral energy
distribution of polarized diffuse Galactic dust emission using a
template-based approach, showing variations between two selected
half-regions. We find that the statistical significance is robust to
the choice of the dust template, \Planck 217 or 353\,GHz, with confidence
levels of $3.2\,\sigma$ and $3.9\,\sigma$ for $\beta_\mathrm{d}$, respectively.
Diffuse dust emission in the \Spider region is not accurately modeled
by a single scaling of a polarized dust template, $\alpha$, over the
entire region.  Future CMB component separation analyses will need to
accommodate the possibility of spatial variation in the SED of diffuse
polarized dust, or quantify the impact of any simplifying
approximations.  If confirmed and interpreted as a spatial variation
in the emissivity index, $\beta_\mathrm{d}$, this would challenge our
understanding of the polarized interstellar medium.

Within a subset of the \Spider region, the joint \Planck 217 and
353\,GHz template analysis provides evidence of a departure from a
simple modified blackbody spectral energy
distribution at more
than $3\,\sigma$, under the assumption that the morphology of dust
emission is consistent at all frequencies within each region. In the
context of a single temperature modified black body considered here, a
break in the spectral index would be required to accommodate this
result.

In Section~\ref{sec:testing_modeling} it is shown that, averaged over
the full region, the spectral energy distribution of polarized dust is
consistent with that of a modified blackbody over the range of
frequencies probed by \Spider and \Planck HFI. The data provide no
significant evidence for variation of the emissivity index with
angular scale. The spatial distribution of the emission is found to be
consistent with a power law in angular multipole. Given that the
highest signal-to-noise dust emission drives the result in the full region,
this finding is consistent with the analysis of the sub-regions.

Component separation may be complicated by line-of-sight frequency decorrelation, such as may result from
multiple dust clouds with differing SEDs and orientations in the same
line of sight. In Section~\ref{sec:los_dust},
we look for evidence of this effect in our region
in locally isolated structures and a global rotation of dust angles
with frequency. We do not find any evidence for this effect.

Looking forward, the addition of 280\,GHz data from the second
flight of \spider in 2022-23 will provide constraints on dust foregrounds in this
region of sky that are both complementary to and independent
of \planck, enabling further exploration of the topics considered
here.

%% file: appendix_orion_superbubble.tex
\section{Impact of Compact Structures on $\alpha_{150}$ Spatial Variation}
\label{app:bubble+MCs}

\begin{figure}
  \centering
  \includegraphics[height=2in]{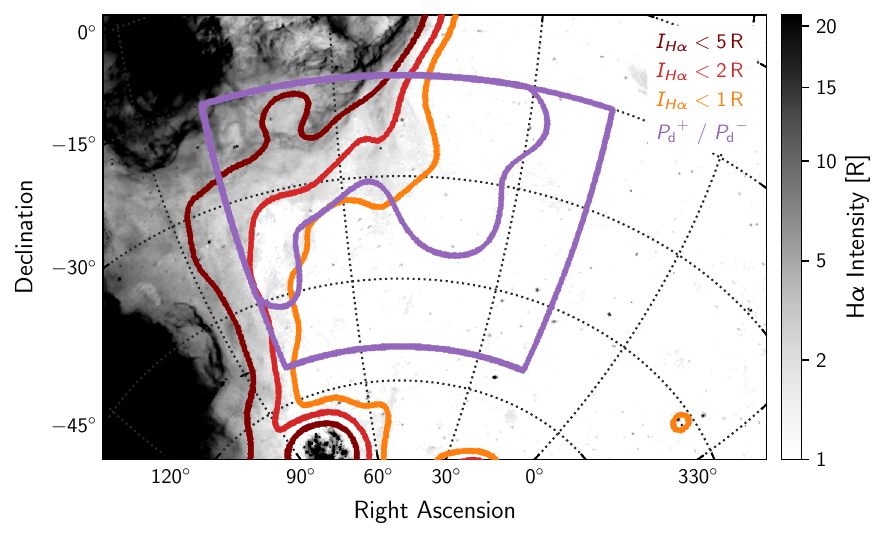}\hfill
  \includegraphics[height=2in]{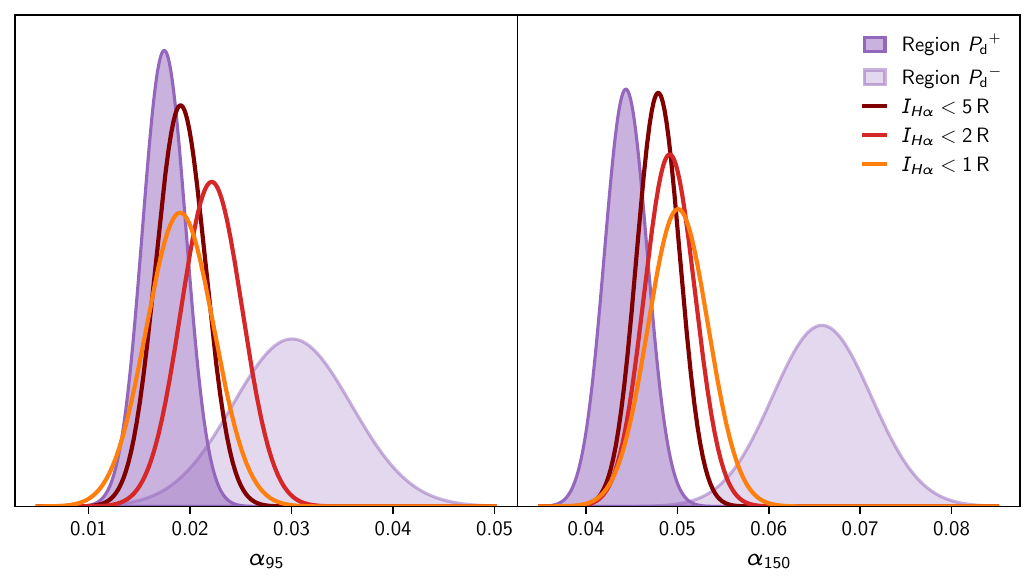}
  \caption{(\textit{left}) The merged all-sky H$\alpha$ map from \citet{Finkbeiner_2003} in
    units of Rayleighs. The map is smoothed and thresholded at several
    intensities to create a series of masks (\textit{red/orange}). The
    \gdust and \ldust subregions of the \spider observing region are also
    overlaid (\textit{purple}).
    (\textit{right}) Plots of $\alpha_{95}$ and $\alpha_{150}$ 1-D likelihoods computed
    with the XFaster framework for the three Orion-Eridanus Superbubble masks,
    and compared to the likelihoods for the \gdust and \ldust regions.
    \label{fig:Halpha}}
\end{figure}

In this Appendix, we assess the impact of the compact structures discussed in Section~\ref{sec:spatial_variation}---the Orion-Eridanus Superbubble and the Magellanic Stream---on the detection of spatial variation in the value of $\alpha_{150}$.

Following \citet{Soler_2018}, we use the full-sky H$\alpha$ map synthesized by \citet{Finkbeiner_2003} as a morphological tracer of the Orion-Eridanus Superbubble in the \Spider region.
In the background map of the left panel of Figure~\ref{fig:Halpha}, the presence of the Orion-Eridanus Superbubble is evident in the top left corner of our region.
To quantify its contribution to the determination of $\alpha_{150}$, we mask the region it occupies with successively more aggressive thresholds.
Specifically, for the reasons discussed in Section~\ref{sec:generating_subregions}, we start by smoothing the \citet{Finkbeiner_2003} map with a $5^\circ$-FWHM Gaussian.
We then build three masks rejecting all pixels in which the H$\alpha$ emission is brighter than 5\,R(ayleighs), 2\,R, and 1\,R, respectively.
The threshold-derived borders of these masks, which leave the \Spider region with, respectively, 4.39\%,  3.81\%, and 3.06\% of the sky, are shown in the left panel of Figure~\ref{fig:Halpha}, along with the border of the \gdust and \ldust subregions ($f_\mathrm{sky}\approx 2.4\%$) used in the spatial variation analysis of Section~\ref{sec:spatial_variation}.
Finally, we compute $\alpha_{95}$ and $\alpha_{150}$ as in Section~\ref{sec:spatial_variation} using the \Planck 353\,GHz template, but for the full \Spider region masked by each of the three H$\alpha$ masks. 
We chose the \Planck 353\,GHz template for this analysis because it provides the dominant contribution to the SMICA reconstructed dust map, as shown in Figure~\ref{fig:smica_all_weights}.
Results from these subregions are shown in the right panel of Figure~\ref{fig:Halpha}.

The value of $\alpha_{95}$ is quite robust to the choice of a Superbubble mask.
In fact, the least aggressive mask results in the same value of $\alpha_{95}$ as the most aggressive mask, a value which remains consistent with that determined in the \gdust region.
Similarly, although the value of $\alpha_{150}$ trends higher as we mask more aggressively---i.e., it trends towards its value in the \ldust region---it remains
largely consistent with its value in the \gdust region, with the \gdust and  $I_{H\alpha} < 1R$ results in agreement at $1.4\,\sigma$.

Therefore, the presence of the Orion-Eridanus Superbubble in the \Spider region does not appear to drive the detected variation in the value of $\alpha_{150}$.

The Magellanic Stream, which falls mostly outside of the \Spider region aside from a few high-velocity clouds (HVCs) along the edge of its bottom right corner, also does not appear to be driving the detected spatial variation in the value of $\alpha_{150}$.
Indeed, none of the H$\alpha$ masks shown in Figure~\ref{fig:Halpha} exclude the HVCs.
Nevertheless, the values of $\alpha_{150}$ derived in the full \Spider region after application of any of these masks are consistent with each other, but inconsistent with those found in the \ldust region, in which the HVCs also remain.
If HVCs were the cause of the detected spatial variation in $\alpha_{150}$, we would expect these four results to be in agreement.

%% file: appendix_dustalpha.tex
\section{Conversion between template $\alpha$ and spectral index $\beta$}
\label{appendix:alpha_beta_conv}
Galactic dust emission is modeled as a single MBB with a power-law emissivity,
\begin{equation}
  I(\nu) = \tau_0 \left( \frac{\nu}{\nu_0} \right)^{\beta} B(\nu, T),
  \label{eq:dust_sed}
\end{equation}
where $\tau_0$ is the optical depth at the reference frequency $\nu_0$.  Throughout the text, we use $\beta_\mathrm{d}$ and $T_\mathrm{d}$ to refer to the spectral index and temperature of the dust component as defined in Equation~\ref{eq:dust_sed}.  We
calculate an effective observing frequency using the dust SED that takes into
account the bandpass of the observing instrument.  In general,
\begin{equation}
\bar{\nu} = \frac{\int d \nu \; \nu I(\nu) t(\nu) }{\int d \nu \; I(\nu) t(\nu) },
\label{eq:nu_eff}
\end{equation}
where $I(\nu)$ is the SED of the observed source and $t(\nu)$ is the spectral
transmission of the detector, properly normalized to its optical efficiency
$\epsilon$. This equation is a generalization of Equation~4 of
\citet{planck2013_spectralresponse}. There, they express $I(\nu)$ for a
power-law~like SED: $I(\nu) \sim \left(\nu/\nu_0
\right)^{\alpha_{Pl}}$.  Note that we identify the spectral index for
this simple SED as $\alpha_{Pl}$ to differentiate it from the template
coefficients $\alpha_{95}$ and $\alpha_{150}$ used throughout the main text.

\begin{table}[t]
\caption{Effective \planck and \spider band centers in GHz for
  different input SED models.  Band center labels correspond to the
  CMB and dust models used throughout the main text.
  \label{table:eff_bandcenter}
}
\hspace{3mm}
\begin{tabular}{rc|cccc|cc}
\toprule
& & \multicolumn{4}{c|}{\Planck} & \multicolumn{2}{c}{\Spider} \\
SED Description & Label & 100\,GHz & 143\,GHz & 217\,GHz & 353\,GHz & 95\,GHz & 150\,GHz \\ \midrule
Flat $I(\nu) = 1$ &  $\bar{\nu}^\mathrm{CMB}$ & 101.2 & 142.6 & 221.4 & 360.6 & 94.8 & 150.8 \\
$\alpha_{Pl} = 4$ &  & 105.1 & 148.1 & 228.8 & 371.4 & 96.9 & 154.0 \\
MBB ($\beta_\mathrm{d} = 1.53, T_\mathrm{d} = 19.6\,\mathrm{K}$) & $\bar{\nu}^\mathrm{d}$  & 104.5 & 147.2 & 227.3 & 368.8 & 96.6 & 153.5 \\
MBB ($\beta_\mathrm{d} = 1.59, T_\mathrm{d} = 22.0\,\mathrm{K}$) & & 104.5 & 147.3 & 227.5 & 369.1 & 96.6 & 153.5 \\
\bottomrule
\end{tabular}
\end{table}

Table~\ref{table:eff_bandcenter} lists effective frequency band centers for a
variety of SED models. The results presented in
\citet{planck2013_spectralresponse} assume the $\alpha_{Pl} = 4$ SED. However, this
assumption holds true only in the Rayleigh-Jeans limit, and only when $\beta
\approx 2$.  A more accurate effective frequency can be calculated by
considering the true dust SED.

CMB maps are typically shown in so-called thermodynamic units ($\mu K_\mathrm{CMB}$),
which encode the amplitude of fluctuations about the CMB temperature,
$T_\mathrm{CMB} = 2.725$\,K. The amplitude of dust emission can be expressed in these units
by normalizing the dust emissivity with a first-order Taylor expansion of the CMB blackbody emission about $T = T_\mathrm{CMB}$:
\begin{equation}
\Delta T (\bar{\nu}^\mathrm{d}) = \left( \left. \frac{\partial B}{\partial T} \right|_{T = T_\mathrm{CMB}} \right)^{-1} I(\bar{\nu}^\mathrm{d})
= \frac{(e^x - 1)^2}{x^2 e^x} \frac{1}{2k} \left(\frac{c}{\bar{\nu}^\mathrm{CMB}}\right)^2  I(\bar{\nu}^\mathrm{d}).
\end{equation}
Here, for convenience, we have defined
$x \equiv h\bar{\nu}^\mathrm{CMB}/(kT_\mathrm{CMB})$. This equation describes
the anisotropy in the mm-map caused by dust emission. It is important to note
that the effective frequency used for conversion to thermodynamic units
($\bar{\nu}^\mathrm{CMB}$) differs from the effective frequency used for the dust
scaling ($\bar{\nu}^\mathrm{d}$). The map-domain scaling relationship from a reference
frequency $\nu_0$ to a lower frequency $\nu$ is then expressed as
\begin{equation}
  A\left(\bar{\nu}^\mathrm{d}, \bar{\nu}_0^\mathrm{d}; \beta_\mathrm{d}, T_\mathrm{d}\right)
  = \frac{\Delta T(\bar{\nu}^\mathrm{d})}{\Delta T(\bar{\nu}_0^\mathrm{d})}
  = \frac{(e^x - 1)^2}{x^2 e^x} \frac{x_0^2 e^{x_0}}{(e^{x_0} - 1)^2}
  \left(\frac{\bar{\nu}_0^\mathrm{CMB}}{\bar{\nu}^\mathrm{CMB}}\right)^2
  \left(\frac{\bar{\nu}^\mathrm{d}}{\bar{\nu}_0^\mathrm{d}}\right)^{\beta_\mathrm{d}}
  \frac{B(\bar{\nu}^\mathrm{d}, T_\mathrm{d})}{B(\bar{\nu}_0^\mathrm{d}, T_\mathrm{d})}.
  \label{eq:big_alpha}
\end{equation}

Because we construct dust templates from \Planck difference maps, specifically
$353-100$\,GHz or $217-100$\,GHz, there is a minor subtraction of dust at
$100$\,GHz. Consequently, the scaling of the difference template to a frequency
$\nu$ is thus given by
\begin{equation}
  \alpha\left(\bar{\nu}^\mathrm{d}, \bar{\nu}_0^\mathrm{d}; \beta_\mathrm{d}, T_\mathrm{d}\right)
  = \frac{\Delta T(\bar{\nu}^\mathrm{d})}{\Delta T(\bar{\nu}_0^\mathrm{d}) - \Delta T (\bar{\nu}_{100}^\mathrm{d})}
  = \frac{A\left(\bar{\nu}^\mathrm{d}, \bar{\nu}_0^\mathrm{d}; \beta_\mathrm{d}, T_\mathrm{d}\right)}
  {1 - A\left(\bar{\nu}_{100}^\mathrm{d}, \bar{\nu}_0^\mathrm{d}; \beta_\mathrm{d}, T_\mathrm{d}\right)},
  \label{eq:beta_to_alpha_conv}
\end{equation}
where $\bar{\nu}_{100}^{d} = 104.5$\,GHz according to Table~\ref{table:eff_bandcenter}.
To recover a value of $\beta_\mathrm{d}$ given $\alpha$,
Equation~\ref{eq:beta_to_alpha_conv} must be inverted using numerical methods.

%% file: appendix_pysm.tex
\section{Comparison to All PySM3 Dust Models}
\label{app:more_pysm}

\begin{figure}
  \centering

  \includegraphics[width=\columnwidth]{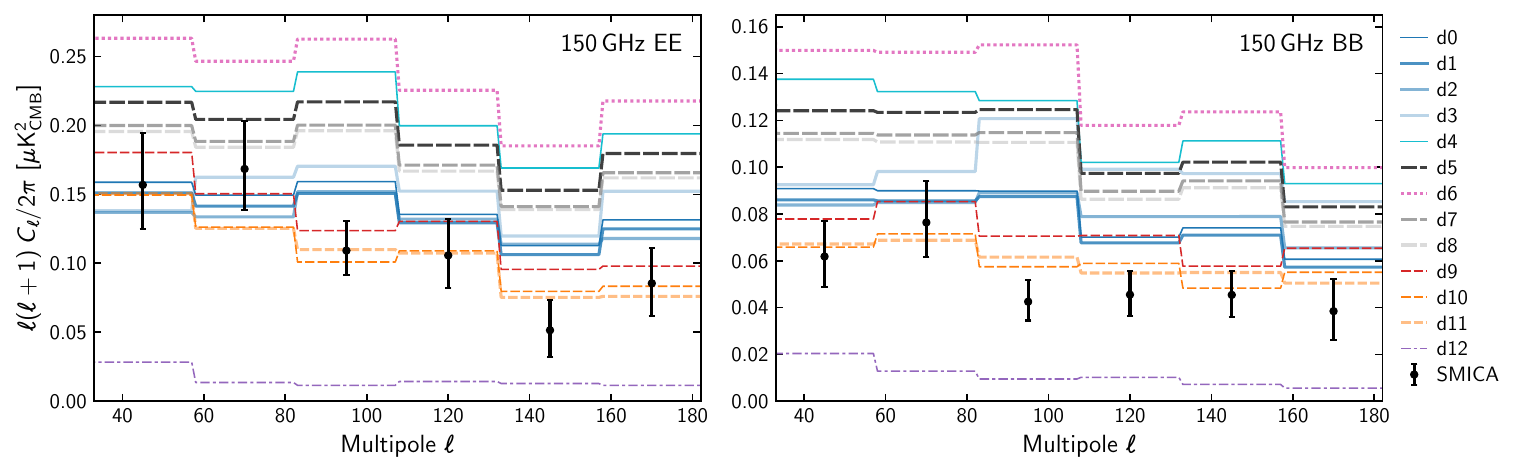}
  \caption{ Companion to Figure~\ref{fig:pysm_comparison} with the
    remaining PySM3 dust models.  As before, this compares the $EE$
    (\textit{left}) and $BB$ (\textit{right}) spectra from the PySM3
    dust models (\textit{colored lines}) to SMICA-derived dust
    bandpowers at 150\,GHz (\textit{black points}) in the \spider
    region.  Models based on the same underlying inputs and class of
    SED modeling are shown with the same line style.
    Models drawn with thinner lines are those already shown in Figure~\ref{fig:pysm_comparison}.
  \label{fig:more_pysm_comparison}}
\end{figure}

Section~\ref{sec:pysm} and Figure~\ref{fig:pysm_comparison} present a comparison between PySM3 dust models and the dust component estimated by SMICA.
For simplicity, that comparison used only a representative subset of the available models.
The remaining ones are shown in Figure~\ref{fig:more_pysm_comparison}.

Due to the fact that both the \Spider SMICA results and the PySM3
models incorporate---to varying extents---both the \Planck data and
noise simulations, the covariance between them is
non-trivial.  While this makes a quantitative measure of the goodness
of fit between the \Spider data and the models difficult to assess, we
reiterate that the overall amplitude of the \Spider SMICA result is
driven by the \Spider 95 and 150 GHz data, and is therefore largely
independent from the amplitude of the models shown.
The amplitudes of the {\bf d9--d11} family of models are in closest agreement with the SMICA amplitudes at 150\,GHz.
Most other models predict more power than \Spider, except {\bf d12}, which predicts significantly less.

For convenience,
we provide a brief description of the PySM3 dust models shown
in Figure~\ref{fig:more_pysm_comparison}.  For a definitive
description we refer the reader to the official documentation\footnote{
   \url{https://pysm3.readthedocs.io/en/latest/models.html\#dust}}.

\input{pysm_model_description}

%% file: pysm_model_description.tex
\begin{description}
\item[Models d0--d4] This family of models treats thermal dust emission as a single-component modified black body (MBB).  Polarized dust templates are based on the \Planck 353\,GHz channel, and are scaled to other frequencies using an MBB spectral energy distribution (SED). The polarization templates are smoothed with a Gaussian kernel of $2.6^\circ$ FWHM, with power at smaller scales added according to the model described in the reference.  For the SED, the models use:\footnote{Here we define families of models by their class designation within PySM3.}
  \begin{description}
  \item[d0] A spatially fixed spectral index $\beta_\mathrm{d}=1.54$ and a black body temperature, $T_\mathrm{d}=20$K.
  \item[d1] A spatially varying spectral index and temperature based upon those provided in the \Planck Commander maps \citep{planck2015_X}, which assume the same spectral index for both intensity and polarization.
  \item[d2 (d3)] An emissivity index that varies spatially on degree scales, drawn from a Gaussian with $\beta=1.59 \pm 0.2~(0.3)$.
  \item[d4] The two component (one hot, one cold) dust model of \cite{finkbeiner1999extrapolation}.
  \end{description}
\item[Models d5, d7, d8] Implementations of the model described in \cite{Hensley_2017}, using:
  \begin{description}
  \item[d5] The baseline model of that reference.
  \item[d7] A model including iron inclusions in the grain composition.
  \item[d8] A simplified version of {\bf d7} in which the interstellar radiation field strength is uniform.
  \end{description}
\item[Model d6] A model that averages over spatially varying dust spectral indices, both unresolved and along the line of sight, using the frequency covariance of \cite{vansyngel_2017} to calculate the frequency dependence.
\item[Models d9--d11] A family of models using a single component MBB with templates based on the \Planck GNILC analysis \citep{Planck_CompSep}, using the 353 GHz color correction described in \citep{Planck_2018_XI}. The frequency modeling uses:
  \begin{description}
  \item[d9] A fixed spectral index $\beta_\mathrm{d}=1.48$ and temperature $T_\mathrm{d}=19.6$K.
  \item[d10] Additional small scale fluctuations in the template, as well as the dust spectral index and temperature.
  \item[d11] Identical to {\bf d10}, but using a different set of seeds and maximum multipole to generate the small scale features.
  \end{description}
\item[Model d12] A three-dimensional model of Galactic dust emission based on \cite{dustMKD}.
\end{description}